\documentclass[aps,pra,reprint,superscriptaddress,floatfix]{revtex4-1}
\pdfoutput=1
\usepackage[T1]{fontenc}
\usepackage[latin9]{inputenc}
\usepackage{color}
\usepackage{units}
\usepackage{textcomp}
\usepackage{amstext}
\usepackage{amssymb}
\usepackage{amsmath}
\usepackage{graphicx}
\usepackage{subscript}
\usepackage{natbib}
\usepackage{float}
\usepackage{wrapfig}
\usepackage[normalem]{ulem} 
\usepackage{upgreek}
\usepackage{lipsum}
\usepackage{longtable}
\begin{document}
\title{Centre line intensity of a supersonic helium beam}

\author{Adri{\`a}  Salvador Palau}
\affiliation{Department of Engineering, Institute for Manufacturing, University of Cambridge, Cambridge, CB3 0FS, UK}

\author{Sabrina D. Eder}
\affiliation{Department of Physics and Technology, University of Bergen, All\'{e}gaten
	55, 5007 Bergen, Norway}

\author{Truls Andersen}
\affiliation{Department of Physics and Technology, University of Bergen, All\'{e}gaten
	55, 5007 Bergen, Norway}

\author{Anders Kom\'{a}r Ravn}
\affiliation{Department of Physics and Technology, University of Bergen, All\'{e}gaten
	55, 5007 Bergen, Norway}

\author{Gianangelo Bracco}
\affiliation{Department of Physics and Technology, University of Bergen, All\'{e}gaten
	55, 5007 Bergen, Norway}
\affiliation{CNR-IMEM, Department of Physics, University of Genova, V Dodecaneso 33, 16146 Genova,
Italy}

\author{Bodil Holst}
\affiliation{Department of Physics and Technology, University of Bergen, All\'{e}gaten
	55, 5007 Bergen, Norway}

\date{\today}

\begin{abstract}
Supersonic helium beams are used in a wide range of applications, for example surface scattering experiments and, most recently, microscopy. The high ionization potential of neutral helium atoms makes it difficult to build efficient detectors. Therefore, it is important to develop beam sources with a high centre line intensity. Several approaches for predicting the centre line intensity exist, with the so-called quitting surface model incorporating the largest amount of physical dependencies in a single analytical equation. However, until now only a limited amount of experimental data has been available. Here we present a comprehensive study where we compare the quitting surface model with an extensive set of experimental data. In the quitting surface model the source is described as a spherical surface from where the particles leave in a molecular flow determined by Maxwell-Boltzmann statistics. We use numerical solutions of the Boltzmann equation to  determine the properties of the expansion. The centre line intensity is then calculated using an analytical integral. This integral can be reduced to two cases, one which assumes a continuously expanding beam until the skimmer aperture, and another which assumes a quitting surface placed before the aperture. We compare the two cases to experimental data with a nozzle diameter of 10 $\upmu\mathrm{m}$, skimmer diameters ranging from 4 $\upmu\mathrm{m}$ to 390 $\upmu\mathrm{m}$,  a source pressure range from 2 to 190 bar, and  nozzle-skimmer distances between 17.3 mm and 5.3 mm. To further support the two analytical approaches, we have also performed equivalent ray tracing simulations. We conclude that the quitting surface model predicts the centre line intensity of helium beams well for skimmers with a diameter larger than $120\;\upmu\mathrm{m}$ when using a continuously expanding beam until the skimmer aperture. For the case of smaller skimmers the trend is correct, but the absolute agreement not as good. We propose several explanations for this, and test the ones that can be implemented analytically. 
\end{abstract}
\maketitle

\section{INTRODUCTION}\label{sec:Introduction}

The supersonic expansion of a gas into vacuum can be used to obtain a molecular beam with high centre line intensities with narrow speed distributions \cite{Campargue1964,Pauly2000,DePonte2006,Scoles1988,Sab_freejet,Even2015}. Such beams are used in different applications, for example surface scattering experiments and atom beam microscopy \cite{JMI:JMI1874,FahyA_2015,ZP_Mue_2,SalvadorPalau2016}.  Noble gas atoms are very hard to detect due to their high ionization potential \cite{Pauly2000}. Therefore, precise prediction of the beam centre line intensity plays an important role in designing instruments and experiments with a sufficient signal to noise ratio.

In a standard supersonic expansion source used in scattering experiments, a pressurised gas expands from a small aperture called a nozzle into a vacuum. The expansion is then collimated using an aperture placed at the end of a conical structure that points towards the nozzle, forming a beam. This conical structure is commonly known as a skimmer (see Fig. \ref{fig:sketch_ellipsoidal}). The problem of precisely determining particle intensities after the skimmer attains different levels of complexity depending on the modified Knudsen number, $\mathrm{Kn^*}$ at the skimmer position, which determines the flow regime close to the skimmer \cite{Bird1976}. \textcolor{black}{The modified Knudsen number was introduced by Bird \cite{Bird1976} to describe the changes in the flow due to backscattering of atoms from the skimmer. }
\begin{equation}\label{eq:mod_knudsen}
\mathrm{Kn^*}=\mathrm{Kn}\left(\frac{2}{5}S_\parallel^2 \right)^{-2/(\eta_p-1)}.
\end{equation}
Where $S_\parallel$ is the parallel speed ratio, a measure of the velocity spread of the beam defined in Sec. \ref{sec:q_s_model}.  $\eta_p$ is the term leading the inverse power law of the repulsive collision model. For a hard sphere gas $\eta_p\rightarrow\infty$, and for the Lennard-Jones potential $\eta_p=13$ \cite{Bird1994}. $\mathrm{Kn}$ is the Knudsen number:
\begin{equation}\label{eq:expansion}
\mathrm{Kn}=\frac{\lambda_0}{r_\mathrm{S}}=\frac{1}{r_\mathrm{S}\sigma\sqrt{2}n},
\end{equation}
where $\lambda_0$ is the mean free path of the gas particles and $r_\mathrm{S}$ is the radius of the skimmer. $n$ is the number density at the skimmer and $\sigma$ is the temperature dependent collision cross section of the gas atoms. In this case, $\sigma$ can be calculated either according to the stagnation temperature, or according to the maximum between the stagnation temperature and the skimmer temperature. For the case of a cold source the collision velocity will be dominated by the warmer skimmer. 
\textcolor{black}{The need for the modified Knudsen number is justified by the change in the mean free path due to backscattering of atoms from the skimmer; In eq. (\ref{eq:expansion}) $\lambda_0$ is the mean free path for particles unaffected by the skimmer presence.}

The Knudsen number is used to estimate the validity of different flow regimes. Navier-Stokes flow can be assumed for  $\mathrm{Kn}<0.2$, and free molecular flow for $\mathrm{Kn}>1$ \cite{Bird1976}. As the gas moves away from the nozzle, the mean free path of the particles increases and therefore the nature of the flow dynamics of the problem changes \cite{Bird1976}. As explained before, we use here the modified Knudsen number, but the discussion of different flow regimes remains the same. The Knudsen number can only be assumed to be smaller than 0.2 in the space very close to the expansion origin (the nozzle), and hence the Navier-Stokes equations can't be generally used to model the flow of the beam close to, and after, the skimmer. Here, Direct Simulation Monte Carlo methods (DSMC), or direct numerical integration of the differential equation (under simplifying assumptions of the physics of the system), can be used to solve the Boltzmann equation \cite{Bird1994,Reisinger2007}.

At $\mathrm{Kn^*}\lesssim 1$, the centre line intensity of the beam is known to be strongly affected by interaction between the beam and particles reflected from the skimmer \cite{Bird1976}. Considering the reflection of particles from the skimmer wall makes solving the Boltzmann equation difficult, as DSMC methods are often computationally heavy. Some work has been done regarding the effect of skimmer geometries  \cite{Bird1976,Hedgeland2005a,braun:3001,Verheijen198463,Hedgeland2005a}. However, much of this work lacks extensive validation due to the lack of experimental data. This, together with the complexity of some of the proposed approaches, has caused some authors to avoid skimmer attenuation by designing experiments where it is not present.

Another relevant contribution to the beam centre line intensity is the exponential decrease of intensity due to free molecular scattering of the beam's atoms with a background gas in the vacuum chambers \cite{Hedgeland2005a,Verheijen198463}. The importance of this contribution will depend on the quality of the pumping system in the experimental set-up and the flux from the nozzle into the expansion chamber.
 
Intensity calculations disregarding both the interaction between the beam and particles reflected from the skimmer, and collisions with background gas were presented in a range of analytical models published in the 1970's and 1980's, based on a Maxwellian velocity distribution of the supersonic expansion \cite{Anderson1965,Beijerinck1981,bossel1974skimming,Sikora1973}. These models coexist with simpler treatments, disregarding the Maxwellian nature of the beam's velocity distribution (usually compensated by including a peak factor), for example \cite{Sab_freejet,Hedgeland2005a,Witham2011,Reisinger2007}.  Others use Beijerinck and Verster 1981 model that incorporates cluster formation and uses the concept of a virtual source \cite{FahyA_2015,Dastoor_paper, Beijerinck1981}. Analytical models have the advantage of requiring only relatively simple numerical solutions of the Boltzmann equation and of directly showing the dependencies with the different variables in the system. Among the most prolific analytical models are various adaptations of the quitting surface model \cite{Sikora1973}.

\textcolor{black}{In the quitting surface model, the spherical quitting surface is assumed to be located at the distance from the nozzle at which the atoms reach molecular flow \cite{Sikora1973}. The atoms then leave the quitting surface following straight trajectories determined by Maxwell-Bolzmann statistics. The ellipsoidal Maxwellian velocity distribution over the surface is given by three parameters: the most probable velocity $\bar{v}$ along the parallel direction (corresponding to the radial direction from the centre of propagation), and the parallel and perpendicular temperatures, respectively $T_{\vert\vert}$ and  $T_\bot$. These two temperatures are associated with the velocity spread of the beam in spherical coordinates \cite{Bird1970}, and in some models are reduced to a simpler description with only a radial temperature $T_{\vert\vert}$ \cite{Sikora1973}.  }

\textcolor{black}{There are two popular ways to estimate the position of the quitting surface: i) calculating the terminal Mach number using the continuum assumption and taking  the position of the quitting surface to be the distance from the nozzle where the terminal Mach number is close to being reached (see for example \cite{Witham2011,asea1}), or ii) directly computing the expansion's temperatures and observing the point where these temperatures de-couple. De-coupling is defined as the point where the perpendicular temperature is much smaller than the parallel temperature. De-coupling is typically assumed at a distance where the temperatures of the expansion fulfil $T_\bot/T_{\vert\vert}\leq 0.01$, thus determining the position of the quitting surface. Alternative cutoff values have also been proposed \cite{Reisinger2007}, providing a certain degree of freedom to the choice of the quitting surface position. Typically, such temperatures are calculated through a numerical solution of the Boltzmann equation. Previous studies already used such an approach to predict the velocity distribution and intensity in the beam expansion \cite{Toennies1977a,Reisinger2007,Pedemonte2003,Usami}. Given that (ii) is more general than (i), we use (ii) in this paper. }

The quitting surface position can either be placed before the skimmer, at the skimmer or after the skimmer. If the quitting surface is taken to be before the skimmer, the parallel temperature $T_{\parallel}$ dominates. This means that the condition $T_\bot/T_{\vert\vert}\leq 0.01$ is reached close to the expansion source, and that the perpendicular temperature of the beam quickly approaches 0. If the quitting surface is calculated to be at or after the skimmer it means that $T_\bot$ tends to 0 slowly. In this case, the perpendicular temperature $T_\bot$ is mostly used in the calculations, and the expansion is assumed to stop at the skimmer, even in the case that its calculation gives a position further away than the skimmer \cite{Sikora1973}. Regardless of where the expansion is assumed to stop, the centre line intensity is then calculated by integrating over the section of the quitting surface seen by the detector through the skimmer.

In this paper, we present a dataset of centre line intensity measurements for a helium atom beam, using several different skimmer apertures and designs, source temperatures, and skimmer to nozzle distances. We benchmark these intensity measurements with the quitting surface model, and discuss its shortcomings. Additionally, we present a ray tracing simulation of the quitting surface model. This is done using a modification of the ray tracing software known as McStas described in detail in \cite{Holst2015}.  This paper contains a large number of variables, many of which are used in several formulas. All formulas are introduced with definitions as they appear in the text. In addition, to make it a bit easier for the reader to keep an overview,  we have included an Appendix E with a table listing all the variables with definitions.   
\section{THEORETICAL FOUNDATION}
\subsection{The supersonic expansion}\label{sec:supersonic_exp}

The expansion of gas through a small nozzle undergoes two different physical regimes: an initial continuum flow, governed by the Navier Stokes equations, followed by a molecular flow regime. In a sonic nozzle (a Laval tube cut-off in the sonic plane), the total flux per unit time (from now on, centre line intensity) stemming from the nozzle is typically calculated using the isentropic nozzle model \cite{Beijerinck1981}. The sonic plane corresponds to the plane where the Mach number $M=v/c=1$ where $v$ is the average velocity of the gas and $c$ the local speed of sound \cite{pauly2000atom}. The equation for the total intensity stemming from a nozzle then reads \cite{Beijerinck1981}:
\begin{equation}\label{eq:intensity_nozzle}
I_0=\frac{P_0}{k_\mathrm{B}T_0}\sqrt{\frac{2k_\mathrm{B}T_0}{m}}\left(\frac{\pi}{4}d_\mathrm{N}^2\right)\sqrt{\frac{\gamma}{\gamma+1}}\left(\frac{2}{\gamma+1} \right)^{1/(\gamma-1)},
\end{equation} 
where $\gamma$ is the ratio of heat capacities ($5/3$ for Helium), and $d_\mathrm{N}$ is the diameter of the nozzle. In theory, this diameter must be corrected with the size of the boundary layer at the nozzle throat. However, this correction can typically be neglected. 
$k_\mathrm{B}$ is the Boltzmann constant, $T_0$ and $P_0$ are the flow stagnation temperature and pressure  \textcolor{black}{inside the nozzle}, $m$ is the mass of a gas particle. In the second flow regime, the expansion of the gas is calculated using the Boltzmann equation, assuming the nozzle is a point source, and using the \textcolor{black}{following collision integral  $\Omega (T_\mathrm{eff})$ (corresponding to the RHS of the Boltzmann equation, that gives the rate of change of molecules in a phase-space element caused by particles that have suffered a collision) \cite{Toennies1977a,Reisinger2007}}. 

\begin{equation}\label{eq:collisionintegral}
\Omega (T_\mathrm{eff})=\left(\frac{k_\mathrm{B}T_\mathrm{eff}}{\pi m} \right)^{\frac{1}{2}}\int_0^\infty Q^{(2)}\left( E\right)\zeta^5 \mathrm{exp}\left(-\zeta^2\right)d\zeta.
\end{equation}
\begin{equation}
\zeta=\sqrt{\frac{E}{k_\mathrm{B}T_\mathrm{eff}}}.
\end{equation}
Where $T_\mathrm{eff}$ is an effective average temperature intermediate to the values of the parallel and perpendicular temperatures, $Q^{(2)}$ is the viscosity cross section and $E$ is the collision energy of two atoms in the centre-of-mass system. For collisions between particles following Bose-Einstein statistics, the viscosity cross section can be written as follows \cite{Holst,Reisinger2007}:
\begin{equation}\label{eq:BOLTZ_viscositycs}
Q^{(2)}(E)=\frac{8\pi\hbar^2}{mE}\sum_{l=0,2,4...}\frac{(l+1)(l+2)}{2l+3}\sin^2(\eta_{l+2}-\eta_l),
\end{equation}
where $\eta_l$ are the phase shifts for orbital angular momentum $l$, obtained solving the scattering of He atoms in the chosen two body potential.

An ellipsoidal Maxwellian velocity distribution is assumed along the whole expansion \cite{Reisinger2007}. The velocity distribution of the atoms in the expansion, $f_\mathrm{ell}$, is defined in spherical coordinates by the two independent temperatures, $T_{\vert\vert}$ and $T_\bot$, and their two corresponding velocities $v_\parallel$ and $v_\bot$ as described in the introduction,
\begin{multline}\label{eq:ellipsoidal_vel_distribution}
f_\mathrm{ell}\left(\vec{v} \right)=n\left( \frac{m}{2\pi k_\mathrm{B}T_{\vert \vert}}\right)^{\frac{1}{2}}\left( \frac{m}{2\pi k_\mathrm{B}T_{\bot}}\right)\cdot
\\
\mathrm{exp}\left( -\frac{m}{2k_\mathrm{B}T_{\vert \vert}}(v_{\vert\vert}-\bar{v})^2-\frac{m}{2k_\mathrm{B}T_{\bot}}v_{\bot}^2 \right).
\end{multline}
The numerical solution of the Boltzmann equation has been implemented for the Lennard-Jones potential (LJ) \cite{Jones1924}, defined as follows:
\begin{equation}
V_\mathrm{LJ}(r_\mathrm{LJ})=4\epsilon\left[\left( \frac{r_\mathrm{m}}{r_\mathrm{LJ}}\right)^{12}-\left(\frac{r_\mathrm{m}}{r_\mathrm{LJ}} \right)^{6}  \right],
\end{equation}
where $r_\mathrm{LJ}$ is the distance between any two interacting particles. $r_\mathrm{m}$ is the distance at which the potential reaches its minimum, for the case of He corresponding to $r_\mathrm{m}=2.974\;\mbox{\normalfont\AA}$, $\epsilon=2.974$ meV \cite{Proceedingspotential}.
A detailed description of the potential and its implementation in the Boltzmann equation can be found in \cite{Reisinger2007}. The simple LJ potential can be replaced by more sophisticated potentials, such as the Tang, Toennies and Yu (TTY) or Hurly Moldover (HM) potentials \cite{PhysRevLett.74.1546,Hurly2000}. However,  results of previous calculations showed that this is only necessary for source temperatures below 80 K \cite{PhysRevA.59.3084,Pedemonte2003,Reisinger2007}. In the present study, the source temperature is higher than 80 K and the LJ potential is adequate.

\textcolor{black}{The numerical solution of the Boltzmann equation in spherical approximation presented here provides the evolution of the gas velocity, and the temperatures  $T_{\vert\vert}$ and $T_\bot$ with respect to the distance from the nozzle.}
\subsection{The quitting surface model}\label{sec:q_s_model}
As mentioned in the introduction, the quitting surface model assumes that the particles leave in molecular flow from a spherical surface of radius $R_\mathrm{F}$ centred at the sonic point. The centre line intensity of the beam is calculated by integrating over all the particles leaving from the quitting surface and arriving at the detector. In 1973, Sikora separated the quitting surface model in two approaches: one corresponding to what he called the \emph{quitting surface model}, and one which he called the \emph{ellipsoidal distribution model}. The first approach assumes a quitting surface placed before the skimmer and a Maxwellian velocity distribution featuring only the radial component of the velocity: $v_\parallel$. The second approach, the ellipsoidal distribution model, assumes an ellipsoidal Maxwellian velocity distribution featuring both $v_\parallel$ and $v_\bot$, together with a quitting surface placed exactly at the skimmer. For the rest of the paper we will refer to the two approaches as Sikora's quitting surface approach and Sikora's ellipsoidal distribution approach.

Sikora's ellipsoidal distribution approach was later adapted by Bossel to be used for expansions stopping before the skimmer. In other words, Sikora's quitting surface approach (assuming a quitting surface placed before the skimmer) was adapted to incorporate ellipsoidal distributions \cite{bossel1974skimming}. To avoid confusion, it is enough to consider the position of the quitting surface itself: in the case of Sikora's ellipsoidal distribution approach, the expansion is considered to stop at the skimmer. In the case of Bossel's approach, the expansion can be chosen to stop at the skimmer or before it. Expansions stopping after the skimmer have thus far not been treated using the quitting surface model. An attempt of doing so is presented in this paper (see Appendix A).

Bossel's approach is the most general approach described so far, as under the right assumptions it reduces to both approaches proposed by Sikora. Bossel's approach corresponds to integrating eq. (\ref{eq:ellipsoidal_vel_distribution}) over the quitting surface area seen by the detector through the skimmer:
\begin{multline}\label{eq:intensity_ellipsoidal}
I_{\mathrm{D}}=\frac{\tau I_0}{2\pi a^2 R_\mathrm{F}^2{L}}\int_0^{r_\mathrm{D}}\int_0^{r_\mathrm{S}}\int_0^{\pi} g(\delta)r\rho\cos^3\beta \epsilon^3 \\ e^{-S^2(1-\epsilon^2\cos^2\theta)}D(b)d\rho dr d\alpha,
\end{multline}
\textcolor{black}{where $r_\mathrm{D}$ is the radius of the detector opening, $r_\mathrm{S}$ is the radius of the skimmer (see Fig. \ref{fig:sketch_ellipsoidal}), and $a$ is the distance between the skimmer and the detector. $r$, $\beta$, $\theta$, $\delta$, $\alpha$ and $\rho$ are geometrical parameters defined in Fig. \ref{fig:sketch_ellipsoidal}. $\tau=\frac{T_{\vert\vert}}{T_\bot}$ is the fraction between parallel and perpendicular temperatures, which is used to simplify the integral through $\epsilon=\left((\tau\sin^2\theta+\cos^2\theta\right)^{-1/2}$. $g(\delta)$ is the angular dependency of the supersonic expansion density at the quitting surface, and $L=\int_0 ^{\frac{\pi}{2}}g(\delta)\sin\delta d\delta$ corresponds to its integral along the quitting surface. $S=\sqrt{\frac{m\bar{v}^2}{2kT_{\parallel}}}$  is the parallel speed ratio at the quitting surface.}

Unfortunately, Bossel's approach has no simple analytical solutions and is often slow to compute over a wide variable space. For $S_i>5$ Sikora showed that both his ellipsoidal distribution approach and  quitting surface approach can be approximated as \cite{Sikora1973}:
\begin{equation}\label{eq:intensity_sikora}
I=I_1\int_0^{2\pi}\frac{d\Phi}{2\pi}[e^{-S_i^2\sin^2\theta_1}]_{\theta_{1\mathrm{max}(\Phi)}}^{\theta_{1\mathrm{min}(\Phi)}}.
\end{equation}
Here, $\Phi$ is the angle of rotation about the beam axis, and $\theta_1$ is the angle between the vector normal to the quitting surface and the vector connecting a given point on the quitting surface with a point in the detector plane. $\theta_{1\mathrm{min}(\Phi)}$ and $\theta_{1\mathrm{max}(\Phi)}$ are the minimum and maximum angles that fulfil the condition that the line connecting a point in the quitting surface and a point in the detector plane must cross the skimmer aperture.  In the case of Sikora's quitting surface approach, $\theta_1$ is defined from a spherical surface of radius $R_\mathrm{F}$, and $S_i=S_\parallel=\sqrt{\frac{m\bar{v}^2}{2kT_{\parallel\infty}}}$ is the parallel speed ratio at the end of the expansion. In the case of Sikora's ellipsoidal distribution approach,  $\theta_1$ is defined from the skimmer aperture \textcolor{black}{(the radius of the quitting surface is then the distance between the nozzle and the skimmer $x_\mathrm{S}$, $R_\mathrm{F}=x_\mathrm{S}$)}, and  $S_i=S_\bot=\sqrt{\frac{m\bar{v}^2}{2kT_\bot}}$ is the perpendicular speed ratio at the skimmer (see Fig. \ref{fig:sketch_ellipsoidal} for a sketch featuring these geometrical terms).

$I_1$ is defined as the intensity arriving at the detector, assuming that there is no skimmer. This can be obtained in two ways:
\[   
I_1 = 
     \begin{cases}
       I_0\pi r_\mathrm{D}^2\eta_\mathrm{D}\frac{1}{(x_\mathrm{S}+a)^2}. &\quad\text{Using eq. (\ref{eq:intensity_nozzle}) for $I_0$}\\
      \eta_\mathrm{D}\pi r_\mathrm{D}^2 n v_{\infty} \left(\frac{x_\mathrm{S}}{x_\mathrm{S}+a}\right)^2. &\quad\text{Using density at skimmer.} \\
     \end{cases}
\]
Here, $\eta_\mathrm{D}$ is the efficiency of the detector in counts/partice. 
Sometimes, one might be interested to obtain the intensity per area. In order to do so, it suffices to divide $I_1$ by $\pi r_\mathrm{D}^2$. 

From eq. (\ref{eq:intensity_sikora}) it can be shown that for $ r_\mathrm{S}\ll x_\mathrm{S}, \; r_\mathrm{S}\ll a$,  $\frac{a}{ r_\mathrm{S}}>>S_i$, and $r_\mathrm{D}<<a$, the intensity arriving at the detector reads \cite{Sikora1973}:
\begin{multline}\label{eq:I_sikora}
I_\mathrm{S}=I_1\Bigg\{ 1-\exp\left[-S_i^2\left(\frac{r_\mathrm{S}(R_\mathrm{F}+a)}{R_\mathrm{F}(R_\mathrm{F}-x_\mathrm{S}+a)}\right)^2\right]\Bigg\},
\end{multline}
$x_\mathrm{S}$ is the distance between the nozzle and the skimmer.
This equation, with the assumption of $S_i=S_\parallel$, and the expansion stopping before the skimmer is usually preferred to using the perpendicular speed ratio, as measuring the parallel speed ratio of atoms is a well established technique \cite{SalvadorPalau2015}. The simplicity of the model has motivated its usage for example to optimize the intensity of helium microscopes \cite{SalvadorPalau2016,PhysRevA.95.013611_ZP}.
\subsection{Scattering contributions}
The atoms leaving the quitting surface do not travel in a perfect vacuum. Rather, they interact with the background gas and the particles scattered from the chamber and skimmer walls. Such interactions can become significant at high nozzle pressures. There have been various approaches for accounting for this, from DSMC simulations, to simpler numerical models based on assumptions on the scattering properties of the skimmer walls \cite{Luria2011, Hedgeland2005a}. \textcolor{black}{Analytical models for the skimmer contributions are so far non-existent due to the difficulty of solving the Boltzmann equation analytically in a typical nozzle-skimmer geometry. The method that has provided a better understanding is the DSMC method (see, for example \cite{Bird1976}). This method is not employed in this paper due to its complexity, but it can be assumed to be the preferable method when precise, localized predictions are desired. }

Here, we choose to only model the interaction with the background gas via free molecular scattering, as it can be modelled by a simple exponential law \cite{Hedgeland2005a,Verheijen198463}:
\begin{equation}\label{eq:scattering_contribution}
\frac{I}{I_\mathrm{S}}= \exp\left(-\sigma^2 n_\mathrm{B_E}x_\mathrm{S}-\sigma^2 n_\mathrm{B_C}a\right).
\end{equation}
$\sigma=\frac{r_\mathrm{m}}{2^{1/2}}$ is the scattering cross section of the atoms in the Lennard-Jones potential. $n_\mathrm{B_E}$ and $n_\mathrm{B_C}$  are the background number densities in the expansion chamber and the subsequent chambers respectively, measured by a pressure gauge placed far away from the beam centre line.  
\subsection{Overall trends}\label{sec:overall}
\textcolor{black}{In this section we qualitatively describe important trends in the expected behaviour of the centre line intensities according to the theory presented above. \begin{enumerate}
\item For skimmers large enough, the exponential term in the equation for centre line intensity becomes negligible, (eq. (\ref{eq:I_sikora})). Thus, increasing the radius of the skimmer further will not lead to an increase in the centre line intensity.
\item Larger skimmers display a decrease in centre line intensity at high pressure. This is due to the fact that a larger skimmer gives a smaller modified Knudsen number (eq. (\ref{eq:mod_knudsen})) for a given pressure. It is known that for smaller modified Knudsen numbers in the so called transition regime, wide angled shock waves can form, which compromise the flow of the beam \cite{Bird1976}. Note that the shock wave behaviour is not modelled by the theory presented above.
\item The closer the skimmer is to the quitting surface ($(R_\mathrm{F}-x_\mathrm{S})\rightarrow 0$); the higher the centre line intensity will be, as the denominator in the exponential in eq. (\ref{eq:I_sikora}) reaches its minimum. This effect is due to the fact that a larger portion of the quitting surface is captured  and this gives a larger centre line intensity.
\item Colder sources produce more intense beams  because the gas passing through the nozzle has a higher density, which ends up influencing the centre line intensity equation (see eq. (\ref{eq:intensity_nozzle})).
\item Numerical solutions of the Boltzmann equation as described in Sec. \ref{sec:supersonic_exp} predict an intensity dip at low source pressures for small skimmers. This dip cannot be extracted from the equations in a simple manner and will be discussed further in the main text.
\end{enumerate}}
\subsection{The ray tracing simulation}
As an independent test of eqs. (\ref{eq:intensity_ellipsoidal}) and (\ref{eq:intensity_sikora}), a ray tracing simulation of the quitting surface expansion was implemented. The simulation was performed using a modification of the ray-trace software package known as McStas described in \cite{Holst2015,doi:10.1080/10448639908233684,Willendrup2004}. 

\begin{figure}[htb]
	\centering
		\includegraphics[width=1\linewidth]{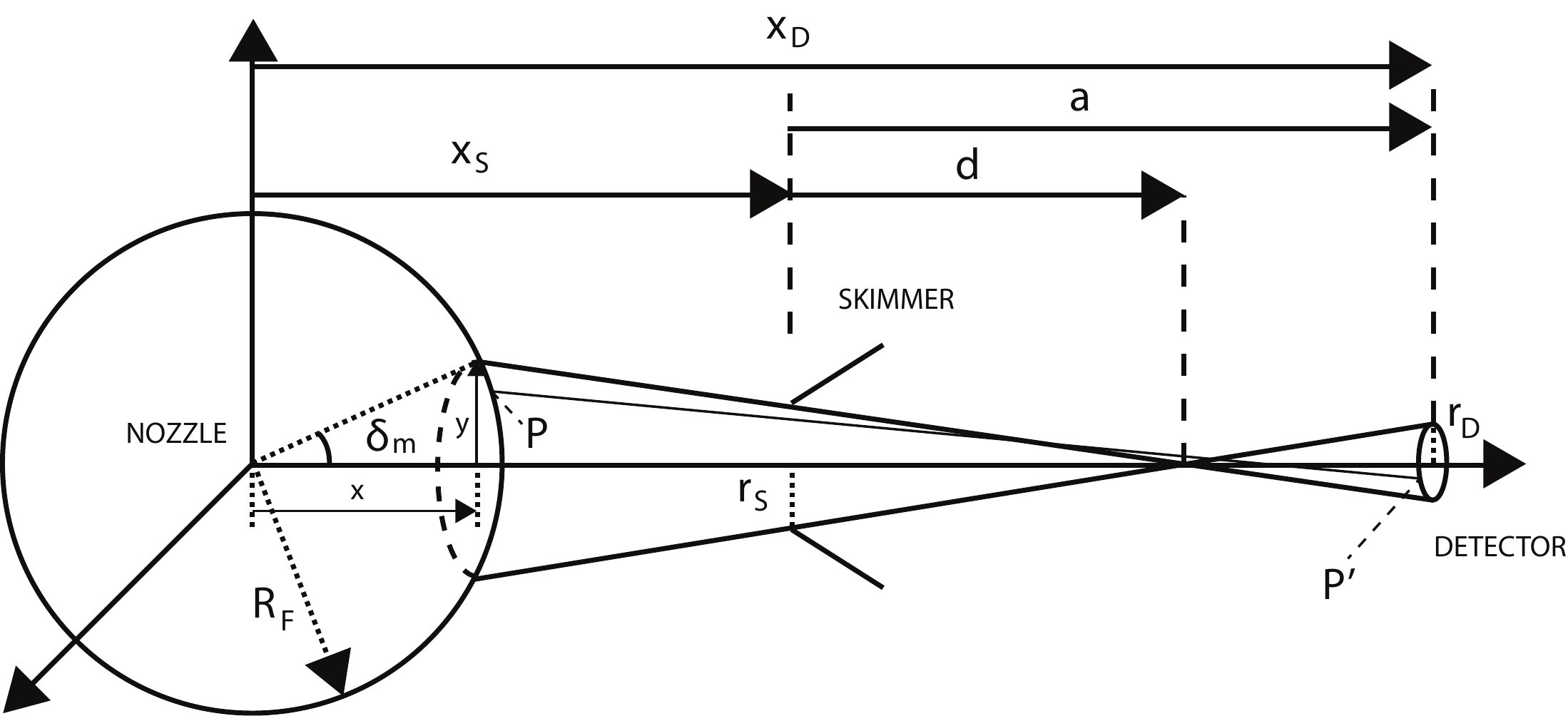}
	\caption{ Diagram of the section of the quitting surface considered in the ray tracing simulation, only the angle $\delta_\mathrm{m}$ seen by the detector through the skimmer contributes to the intensity at the detector. $R_\mathrm{F}$ is the radius of the quitting surface, $y$ is the distance between the axis of symmetry and the projection of the maximum-angle ray on the quitting surface, $r_\mathrm{S}$ is the skimmer radius and $r_\mathrm{D}$ is the radius of the detector. $a$ is the distance between the skimmer and the detector, $d$ is the distance from the skimmer to the point where the maximum-angle ray crosses the symmetry axis. $x_\mathrm{D}$ is the distance between the nozzle and the detector and $x_\mathrm{S}$ is the distance between the nozzle and the skimmer. $x$ is the distance from the point of emission of the maximum angle ray to the nozzle plane. \label{fig:LIMIT_ANGLE}}
\end{figure}

In order to replicate the dynamics assumed during the derivation of eq. (\ref{eq:intensity_ellipsoidal}), a spherical source with ellipsoidal Maxwellian velocity distributions and an anisotropic number density was programmed. \textcolor{black}{The McStas software works with sources featuring uniform spatial ray probability distributions that are later corrected for their real probability weights determined by the physics of the system (in this case, the Maxwellian velocity distribution of the source, and the anisotropic number density)}. This poses a problem when simulating the quitting surface because most of the rays yield probabilities that are too low, bringing insufficient sampling at the detector. To avoid this effect, we only computed the particles stemming from the surface of the quitting surface seen by the detector through the skimmer (see Fig. \ref{fig:LIMIT_ANGLE}). This reduces the computation power needed for each experiment and therefore allows for better statistics in the detector.

\begin{figure}[H]
  \centering
    \includegraphics[width=0.48\textwidth]{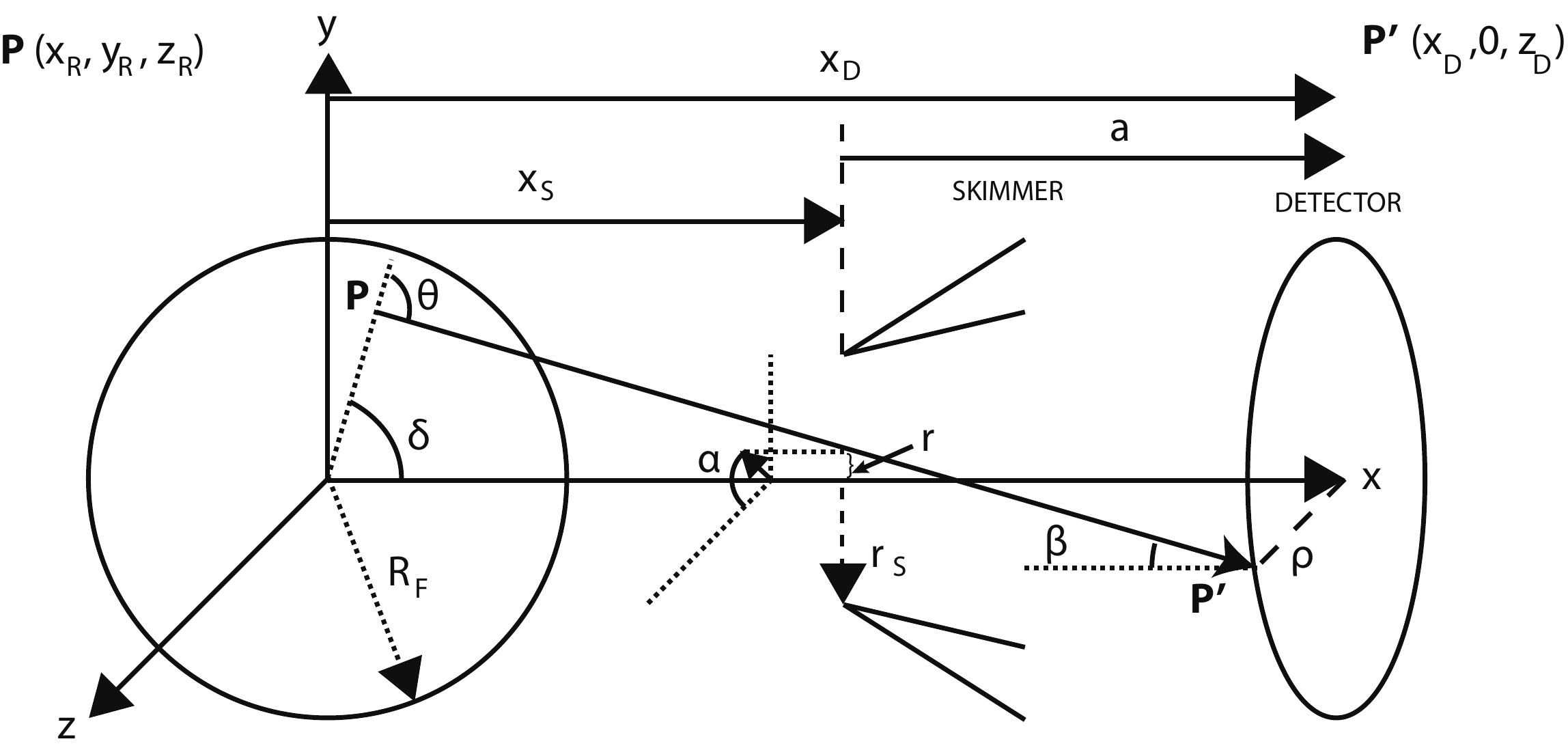}
  \caption{Illustration of all variables used in the ellipsoidal quitting surface model. \textbf{P} is a point on the quitting surface from which a particle leaves in a straight trajectory until \textbf{P'}, a point placed on the detector plane. The point on the quitting surface is given by the set of Cartesian coordinates $(x,y,z)$, which can be related to the polar coordinates $r,\alpha,\rho$ for integration. $x_\mathrm{S}$ is the distance from the nozzle to the skimmer and $x_\mathrm{D}$ is the distance from the nozzle to the detector. Therefore $a=x_\mathrm{D}-x_\mathrm{S}$. The angles $\beta$ and $\theta$ can also be expressed in terms of $r$, $\alpha$ and $\rho$.}
  \label{fig:sketch_ellipsoidal}
\end{figure}

The simulation is performed as follows: first, a circular target or focus of interest is set, which determines the area of the detector, where the rays will hit. Then, the point P' is generated randomly over the area of the detector. Subsequently, a point P over the quitting surface is randomly generated and its connecting vector $\vec{r}$ is computed. Only the points visible by the detector through the skimmer are allowed (see Fig.  \ref{fig:LIMIT_ANGLE}). Therefore a maximal angle $\delta_m$ is set (see the derivation in Appendix C).
\begin{multline}
\delta_m=\arcsin\frac{y}{R_\mathrm{F}}\\ =\arcsin\left( \frac{\frac{d}{r_\mathrm{S}}(d+x_\mathrm{S})-\sqrt{\frac{d^2}{r_\mathrm{S}^2}R_\mathrm{F}^2+R_\mathrm{F}^2-(d+x_\mathrm{S})^2}}{R_\mathrm{F}(\frac{d}{r_\mathrm{S}})^2+R_\mathrm{F}}\right)
\end{multline}
With $d$ corresponding to the distance from the skimmer to the point where the maximum-angle ray crosses the symmetry axis (see Fig. \ref{fig:LIMIT_ANGLE}):
\begin{equation}
d=\frac{a r_\mathrm{S}}{r_\mathrm{D}+r_\mathrm{S}}.
\end{equation}
Which means that the point P must be contained within the following angles:
\begin{equation}
\delta = (0,\delta_m),\; \phi=(0,2\pi).
\end{equation}
In Cartesian coordinates, P is:
\begin{equation}
P=R_\mathrm{F}\left(\sin\delta \cos\phi, \sin\delta \sin\phi,\cos\delta\right).
\end{equation}
Following, a scalar velocity $v$ is randomly generated between two limiting values along the direction of the vector $\vec{r}$. From its Cartesian components, the perpendicular and parallel velocities are obtained:
\begin{multline}
v_{\vert\vert}=\vec{v}\cdot\vec{u}_r=v_x\sin\delta\cos\phi+v_y\sin\delta\sin\phi+v_z\cos\delta,\\
v_{\bot}=\vec{v}\cdot\vec{u}_\delta=v_x\cos\delta\cos\phi+v_y\cos\delta\sin\phi-v_z\sin\delta,\\
v_{\bot'}=\vec{v}\cdot\vec{u}_\phi=-v_x\sin\phi+v_y\cos\phi.
\end{multline}
A probability weight factor given by the Maxwellian velocity distribution of the beam is set for the  ray travelling from P to P' (see Figs. \ref{fig:LIMIT_ANGLE} and \ref{fig:sketch_ellipsoidal}). The intensity recorded at the detector will be the sum of all probability weight factors. Therefore, we can recover eq. (\ref{eq:dif_N}) (Appendix B) in angular coordinates to infer the intensity contributions:
\begin{equation}
dI = \frac{I_0 A_\mathrm{D}}{A_\mathrm{S}L}f_\mathrm{ell}(\vec{v})g(\delta) v^2d\Omega dv.
\end{equation}
$A_\mathrm{D}=\pi r_\mathrm{D}^2$ is the area of the detector. For the experiments presented here, this corresponds to the area of the pinhole placed in front of the detector (see Fig. \ref{fig:exp_setup}), $A_\mathrm{S}\approx\pi y^2$ is the area of the section of the sphere from which particles are simulated assuming $r_\mathrm{S}\ll R_\mathrm{F}$ (the computed section of the quitting surface is small enough relative to $R_\mathrm{F}$ that its area approximates to the area of a circle).  $L$ is defined as in eq. (\ref{eq:L_definition}) (Appendix B) but taking care to integrate only between 0 and $\delta_m$. $d\Omega$ is the solid angle seen through the skimmer from the centre of the detector, this is approximately the same as the solid angle seen from P' through the skimmer. This approximation is true for detectors placed sufficiently far away from the skimmer.
\section{EXPERIMENTAL SETUP FOR INTENSITY MEASUREMENTS}\label{sec:expmethod}
\begin{figure*}[htb]
	\centering
		\includegraphics[width=1\linewidth]{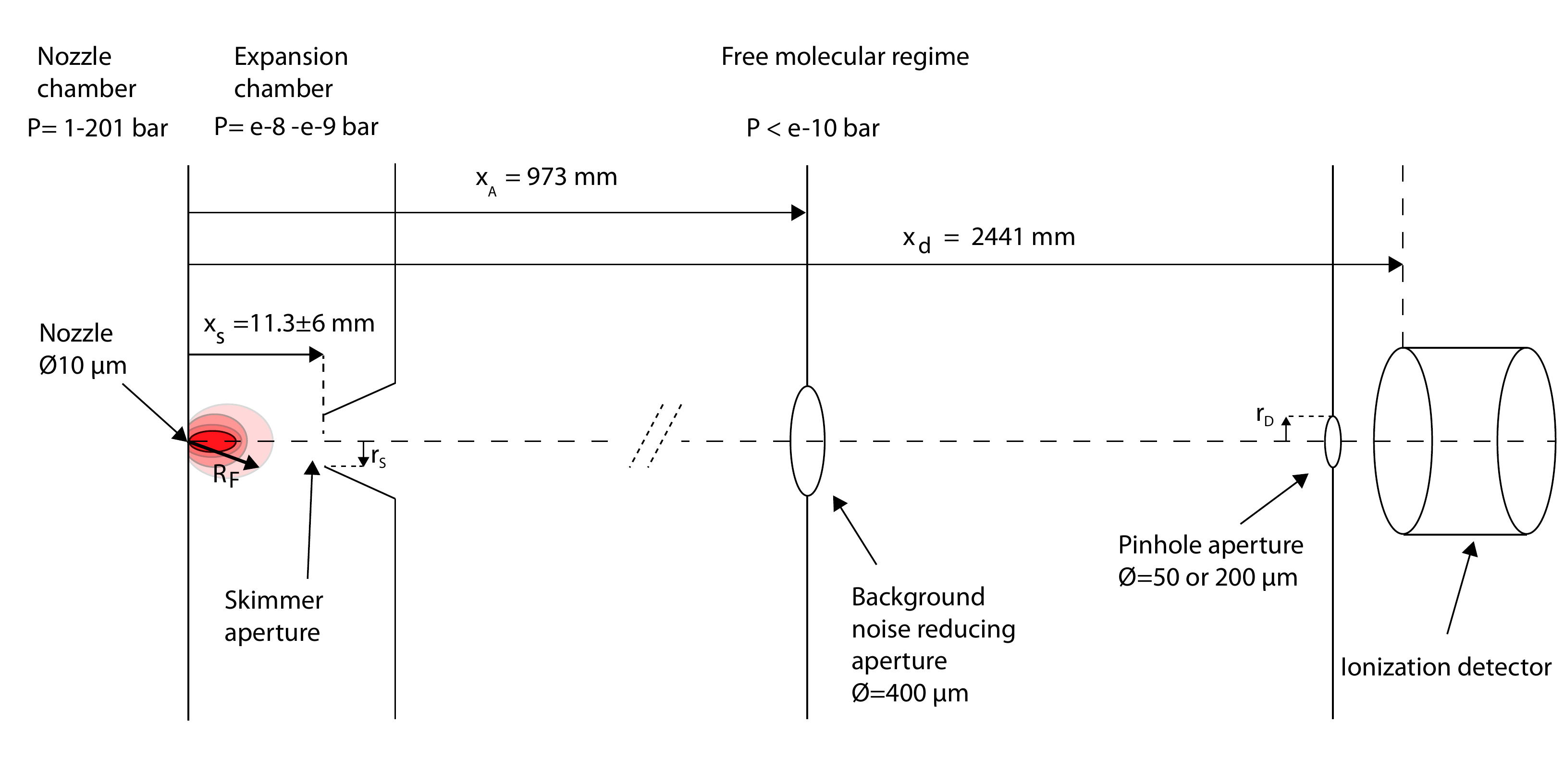}
	\caption{Sketch of the experimental setup used for the centre line intensity measurements. A skimmer is used to select the supersonic beam, followed by two apertures. Vacuum pumps are placed in each chamber to reduce interactions of reflected particles with the beam. $R_\mathrm{F}$ is the radius of the quitting surface, from where the gas particles are assumed to leave following a mollecular flow. \label{fig:exp_setup} }
\end{figure*}

\begin{figure}[htb]
	\centering
		\includegraphics[width=1\linewidth]{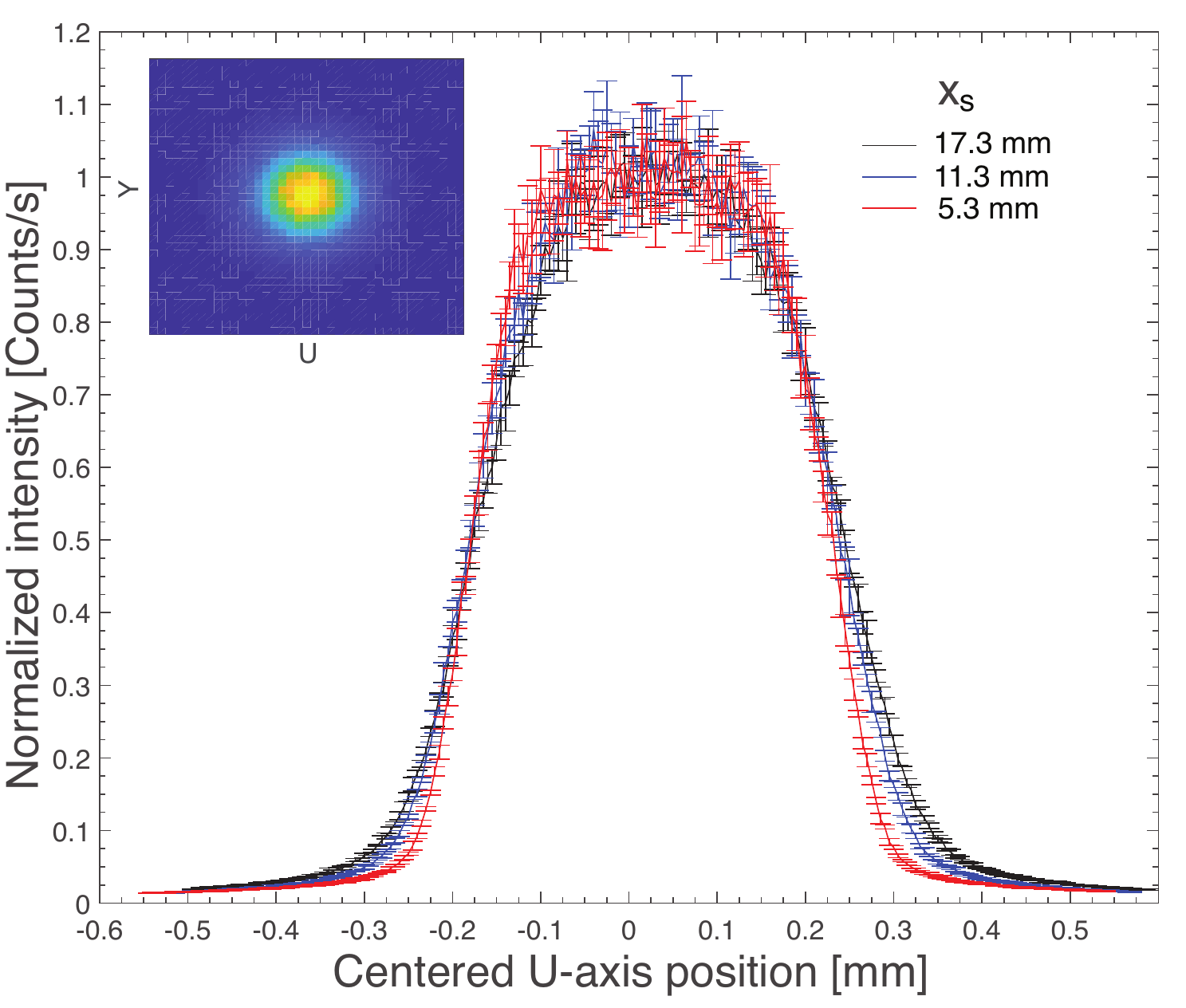}
	\caption{Example of the alignment procedure, here done for a cold source at 60 bar and a 390 $\upmu\mathrm{m}$ diameter skimmer. The nozzle is moved relative to the skimmer in 50 nanometer steps for three values of $x_\mathrm{S}$ \label{fig:alig_source}. The optimum alignment position of  the nozzle relative to the skimmer is obtained by finding the centre point of the maximum of intensity. The complete intensity plot of the beam is shown in the upper left corner. }
\end{figure}

The setup used to obtain the experimental measurements presented in this paper is shown in Fig. \ref{fig:exp_setup}. All the measurements have been obtained using the molecular beam instrument at the University of Bergen, known as MAGIE. This instrument is equipped with a home-built source which enables the skimmer and nozzle to be positioned relative to each other with 50 nm precision \cite{Sab_freejet}. This is particularly important to ensure proper alignment in centre line intensity experiments using small skimmers. A detailed description of the system can be found in \cite{Apfolter2005}. In contrast to most other helium atom scattering instruments with time-of-flight detection, MAGIE has a movable detector arm, which allows us to measure the straight through intensity of the beam without any sample. A centre line intensity measurement is performed by setting the initial pressure in the inlet channel and measuring the inlet channel temperature. For the experiments presented here, the beam source is either "warm" (at ambient temperature) or "cold" (at roughly 125 K). The helium gas expands through a pinhole aperture nozzle, $10\; \upmu \mathrm{m}$ in diameter to a lower pressure chamber where it undergoes a supersonic expansion. We use a Pt-Ir electron microscope aperture as nozzle (purchased
from Plano GmbH, A0301P) \cite{Sab_freejet}. The expansion is then collimated by a skimmer placed $5.3\pm 0.1\; \mathrm{mm}$, or $11.3\pm 0.1\; \mathrm{mm}$, or $17.3\pm 0.1\; \mathrm{mm}$ away from the nozzle. Figure \ref{fig:alig_source} shows an example of the alignment procedure. The nozzle is moved across the skimmer opening in $50~\mathrm{nm}$ steps in a 2D array and eventually moved to the position of maximum intensity which is clearly visible. Note that a displacement of just $0.2~\mathrm{mm}$ leads to a noticeable change in intensity.

Further downstream, at 973 mm from the nozzle, a $400~\upmu \mathrm{m}$ aperture is placed to further reduce the background pressure and thus minimize the beam attenuation. Finally, at $2441~ \mathrm{mm}$ from the nozzle an ionization detector is set. The detector has an efficiency of $\eta_\mathrm{D}=2.1\cdot 10^{-6}$ (provided by the manufacturer). Just in front of the detector another aperture is placed. Two different apertures with diameters $200~\upmu\mathrm{m}$ and $50~\upmu\mathrm{m}$ respectively, were used in the experiments. This allows us to measure the centre line intensity. A table with the diameter of the aperture for each intensity experiment is given in Appendix D.

\begin{figure*}[htb]
	\centering
		\includegraphics[width=0.9\linewidth]{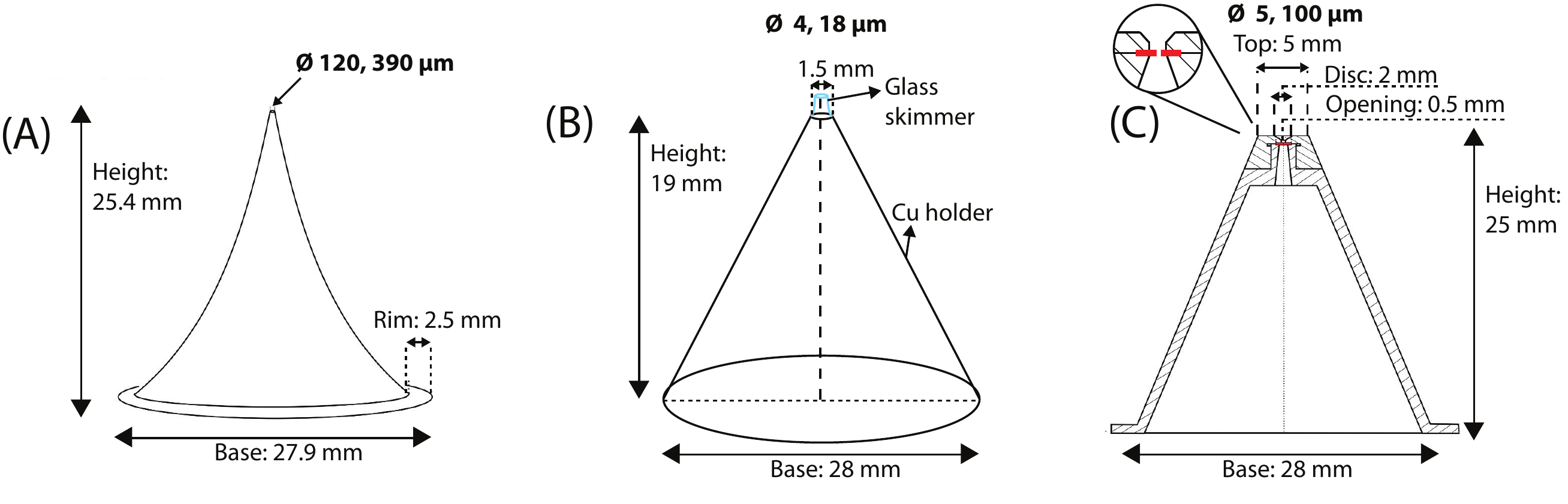}
	\caption{Drawings of the skimmers used for the centre line intensity measurements. 
	(A) corresponds to the Beam Dynamics skimmers, with diameters of 120 and 390 $\upmu\mathrm{m}$, (B) to glass micro-skimmers mounted on copper, with diameters of 4 and 18 $\upmu\mathrm{m}$ and (C) corresponds to the Kurt skimmer, with inserted appertures of 5 and 100 $\upmu\mathrm{m}$ \label{fig:skimmers}.  }
\end{figure*}

Five skimmers were used to collimate the beam, two made of nickel, two made of glass and an additional metallic skimmer known as the Kurt skimmer. The nickel skimmers have apertures  $120$ and $390$ $\upmu \mathrm{m}$ in diameter. They are produced by Beam Dynamics (model 2) and have a streamlined profile \cite{Beamdynamics} (see dimensions in Fig. \ref{fig:skimmers}). The glass skimmers are home made using a \emph{Narishige} PP-830 glass pulling machine, using Corning 8161 Thin Wall capillaries with an outer diameter of 1.5 mm and an inner diameter of 1.1 mm. The glass skimmers are mounted on a Cu holder (see dimensions in Fig. \ref{fig:skimmers}). Their apertures are 18 and 4 $\upmu \mathrm{m}$ respectively, measured using an electron beam microscope. \textcolor{black}{Stereo microscope measurements on the glass skimmers showed an outer opening angle of $\approx32.5^{\circ}$ for the first 200$\upmu\mathrm{m}$, followed by a more narrow section of $\approx12.5^{\circ}$. The inner opening angle could not be determined, but due to the thin opening lip ($\approx 200\mathrm{nm}$), it is expected to be similar to the outer opening angle. This corresponds to what is known as a \emph{slender} skimmer. Slender skimmers are known to produce better performance than wide angle skimmers, as long as the modified Knudsen number at the skimmer is kept large enough \cite{Bird1976}. This condition is fulfilled in the experiments presented here due to the large values of $S_{\vert\vert}$ and the small skimmer openings.  }

The Kurt skimmer is also home made. It is designed to be used with interchangeable apertures on 2 mm diameter discs. Two apertures are used in this study: 5 and 100 $\upmu\mathrm{m}$ in diameter. The dimensions of the Kurt skimmer can be found in Fig. \ref{fig:skimmers} (note the inverted cone shape before the aperture). The Kurt skimmer is made of stainless steel type 1.4301.
\section{RESULTS}\label{sec:Res}
Throughout Figs. \ref{fig:warmall}-\ref{fig:Kurt_effect}  we use open circles for the nozzle-skimmer distance $x_\mathrm{S}$= 5.3 mm, triangles for $x_\mathrm{S}$= 11.3 mm, and asterisks for $x_\mathrm{S}$= 17.3 mm. The labels are included in Fig. \ref{fig:warmall} only. Error bars are not included in the plots because they are too small to show.
\subsection{Ray tracing benchmarking of the centre line intensity integral}
\begin{figure}[htb]
	\centering
		\includegraphics[width=1\linewidth]{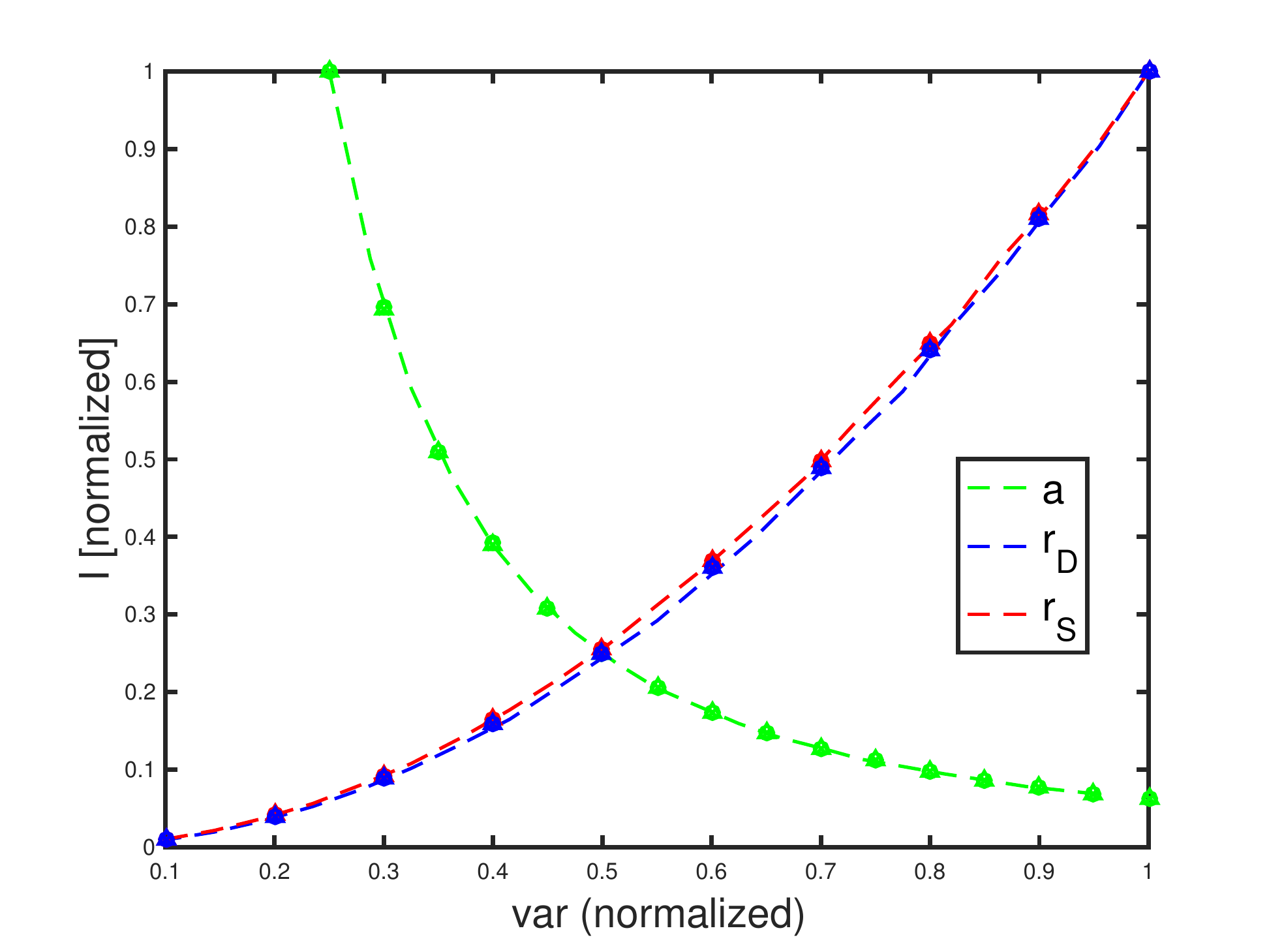}
	\caption{Plot of the ray tracing simulation (dashed lines) compared with eq. (\ref{eq:intensity_ellipsoidal}) and (\ref{eq:I_sikora}) (respectively, circles, crosses for $S_i=S_\parallel$ and triangles for $S_i=S_\bot$, superposed). The green line show the effect on the centre line intensity of varying the distance between the skimmer and the detector, $a$. The blue and red lines show the intensity change when varying the radius of the pinhole in front of the detector, $r_\mathrm{D}$, and the radius of the skimmer, $r_\mathrm{S}$. The centre line intensity and the variable values have been normalized to 1 in order to show all dependences in a single plot. The calculations are done at a fixed skimmer position $x_\mathrm{S}=11.3$ mm (the centre position). $a$ is varied between 0.5 m and 2 m, $r_\mathrm{D}$ is varied between 10 $\upmu\mathrm{m}$ and 100 $\upmu\mathrm{m}$, and the radius of the skimmer, $r_\mathrm{S}$ is varied between 1 $\upmu\mathrm{m}$ and 10 $\upmu\mathrm{m}$. While a variable is varied, the others are kept fix at the maximum value of their span ($a=2$ m, $r_\mathrm{D}=100$ $\upmu\mathrm{m}$, $r_\mathrm{S}=10$ $\upmu\mathrm{m}$. The source temperature is 115 K and the source pressure is 161 bar. Both the ray tracing simulation and the centre line intensity model assume a quitting surface placed just before the skimmer position ($R_\mathrm{F}=11.2$ mm).\label{fig:Monte Carlo_vs_qs}}
\end{figure}

A spherical quitting surface is simulated using the ellipsoidal quitting surface velocity distribution defined in eq. (\ref{eq:ellipsoidal_vel_distribution}). The centre line intensity obtained through the ray tracing simulation is then compared with eqs. (\ref{eq:intensity_ellipsoidal}) and (\ref{eq:I_sikora}) for different spans of the different variables present in the equation. In all cases the result from the analytical models lies within the statistical margin of error of the simulation (see Fig. \ref{fig:Monte Carlo_vs_qs}). In the further sections of this paper we will just show the results from eqs. (\ref{eq:intensity_ellipsoidal}) and (\ref{eq:I_sikora}).

\subsection{120 $\upmu\mathrm{m}$ and 390 $\upmu\mathrm{m}$ skimmers}
In this section, the measured intensities for the large skimmers from Beam Dynamics (see Fig. \ref{fig:skimmers}) (120 and 390 $\upmu\mathrm{m}$ diameters),  are compared with the predictions from eq. (\ref{eq:scattering_contribution}) for the two variations of the model described in Sec. \ref{sec:supersonic_exp}. 
\subsubsection{Warm source, $T_0\approx 300K$}
\begin{figure}[htb]
	\centering
		\includegraphics[width=1\linewidth]{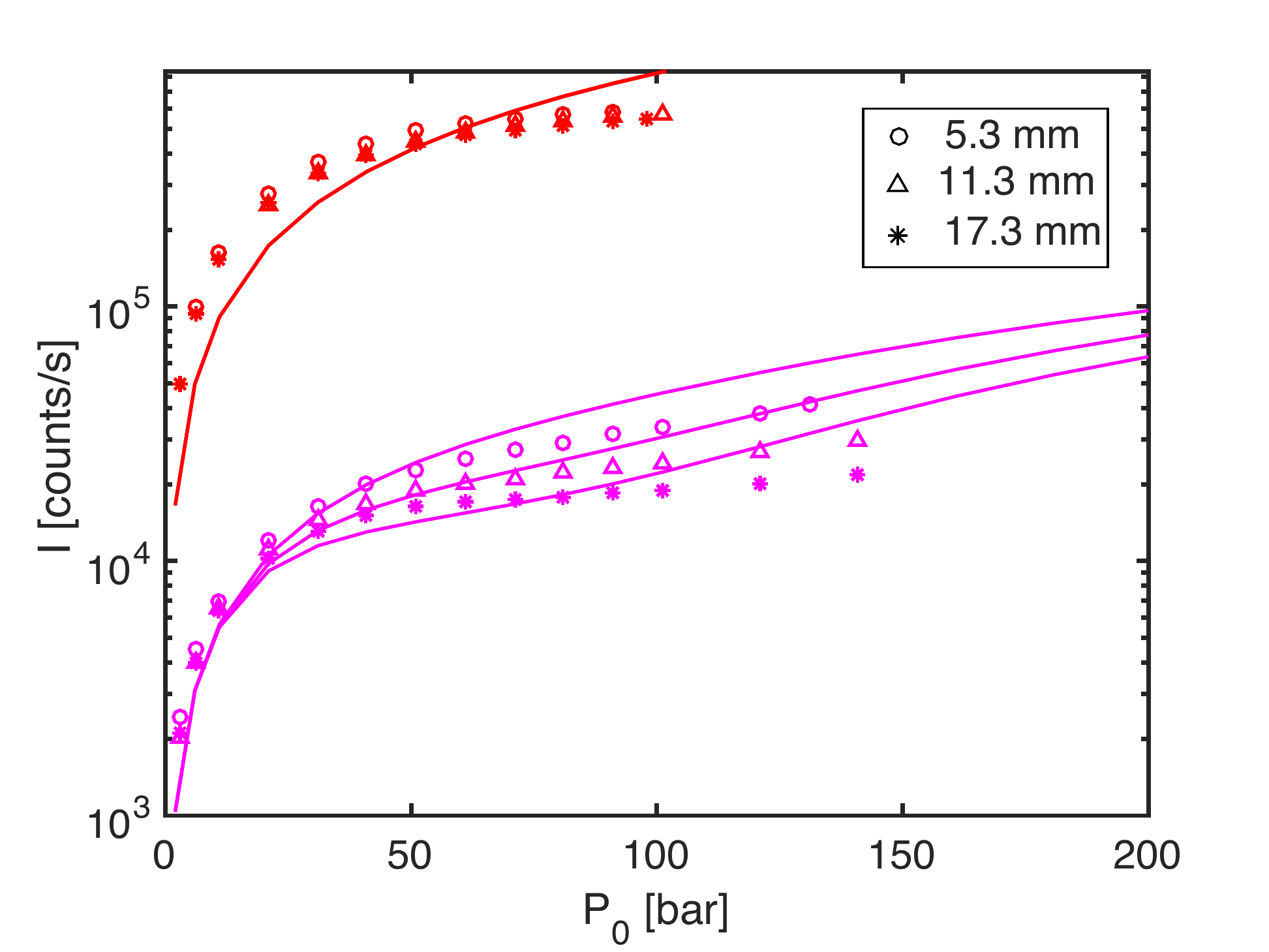}
	\caption{Plot of measured and predicted intensities for a warm source (300 K), 120$\upmu\mathrm{m}$ (pink) and 390$\upmu\mathrm{m}$ (red) skimmers, and for three values of $x_\mathrm{S}$: 5.3 mm (circles), 11.3 mm (upwards arrows) and 17.3 mm (asterisks). The intensities are computed assuming that the expansion stops at the skimmer with $S_i=S_\bot$ . Note that, for the larger skimmer, the centre line intensity becomes independent of the distance between the skimmer and the nozzle, so that all the curves collapse in one simulated curve (in good agreement with what is observed experimentally). The difference in intensities between the two skimmers is due to the fact that they were obtained using different pinholes in front of the detector (see Appendix D and Fig. \ref{fig:exp_setup}) \label{fig:warmall}}
\end{figure}

\begin{figure}[htb]
	\centering
		\includegraphics[width=1\linewidth]{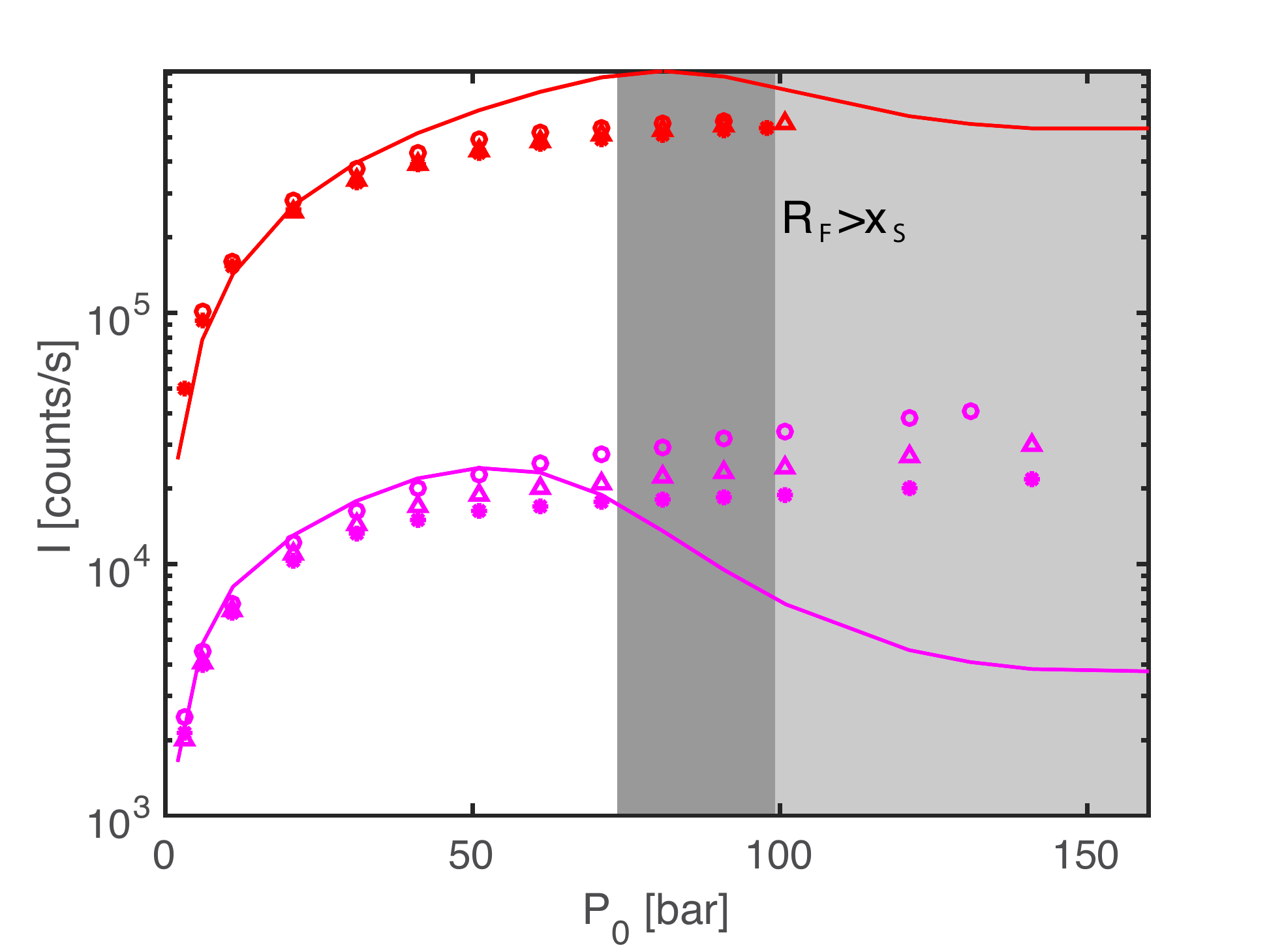}
	\caption{Plot of measured intensities for a warm source (300 K), and 120$\upmu\mathrm{m}$ (pink) and 390$\upmu\mathrm{m}$ Beam Dynamics skimmers (red). The measured intensities are compared to eq. (\ref{eq:scattering_contribution}), with the expansion stopped before the skimmer and $S_i=S_\parallel$. Note how after the quitting surface has surpassed the skimmer, the model loses its predictability (light grey for $x_\mathrm{S}=17.3\; \mathrm{mm}$, dark grey indicates the whole span for the different values of $x_\mathrm{S}$ ). The difference in intensities between the two skimmers is due to the fact that they were obtained using different pinholes in front of the detector (see Appendix D). \label{fig:warmQS_QS}}
\end{figure}

The results for a warm source are shown in Figs. \ref{fig:warmall} and \ref{fig:warmQS_QS}. Fig. \ref{fig:warmall}  shows the experimental results and eq. (\ref{eq:scattering_contribution}) with the  expansion assumed to stop at the skimmer, and $S_i=S_\perp$. The experimental results are reproduced fairly well over the whole range, but with a trend towards too high theoretical values for higher pressures. To obtain $n_\mathrm{B_E}\rightarrow n_\mathrm{B_E}(P_0)$ for eq. (\ref{eq:scattering_contribution}), we use a set of measured background pressures in the expansion chamber. From observation this dependency is linear, and the equation obtained is:
\begin{equation}\label{eq:densities}
n_\mathrm{B_E}=\frac{1}{k_\mathrm{B}T_0}\left(m_E\cdot P_0+n_E\right).
\end{equation}
$m_E$ and $n_E$ are the linear fit coefficients from fitting the measured background pressures $P_B$ with respect to $P_0$. Concretely, for this set of measurements $m_E=3.9\cdot 10^{-4} \frac{Pa}{bar}$, $n_E=-5.8\cdot 10^{-4} Pa$ if $P_0$ is given in bar and $n_\mathrm{B_E}$ in SI units (positive values of $n_\mathrm{B_E}$ are guaranteed by the experimental pressure range, $P_0\geq 2$ bar). The number density after the skimmer, $n_\mathrm{B_C}$, was experimentally measured to be approximately 1/20 of $n_\mathrm{B_E}$, eq. (\ref{eq:densities}) was used with the corresponding factor.

Fig.  \ref{fig:warmQS_QS} shows the values of eq. (\ref{eq:scattering_contribution}) for the 120 $\upmu\mathrm{m}$ and 390 $\upmu\mathrm{m}$ skimmers, where the expansion is assumed to stop before the skimmer (in this case for $T_\bot/T_\parallel\leq 0.1$), and $S_i=S_\parallel$. At small source pressures there is good agreement between experiments and simulations, but the dependency on the nozzle-skimmer distance is lost. At high pressures the model becomes non-physical because the point at which $T_\bot/T_\parallel\leq 0.1$ is calculated to be positioned after the skimmer. One must note that the decrease in centre line intensity at high pressures is not given by the model (eq. (\ref{eq:I_sikora})) being un-physical, but instead by $S_\parallel^2 r_\mathrm{S}^2/R_\mathrm{F}^2\rightarrow 0$ as $P_0$ increases. If the expansion is assumed to always stop at the skimmer ($R_\mathrm{F}=x_\mathrm{S}$) as in the case of Fig. \ref{fig:warmall}, this condition does not hold any more and the predicted centre line intensity increases monotonically with $P_0$. In this case, eq. (\ref{eq:scattering_contribution}) is also used. The discrepancy at low pressures is discussed in Sec. \ref{sec:discussion}.


\subsubsection{Cold source $T_0\approx 125 K$}
\begin{figure}[htb]
	\centering
		\includegraphics[width=1\linewidth]{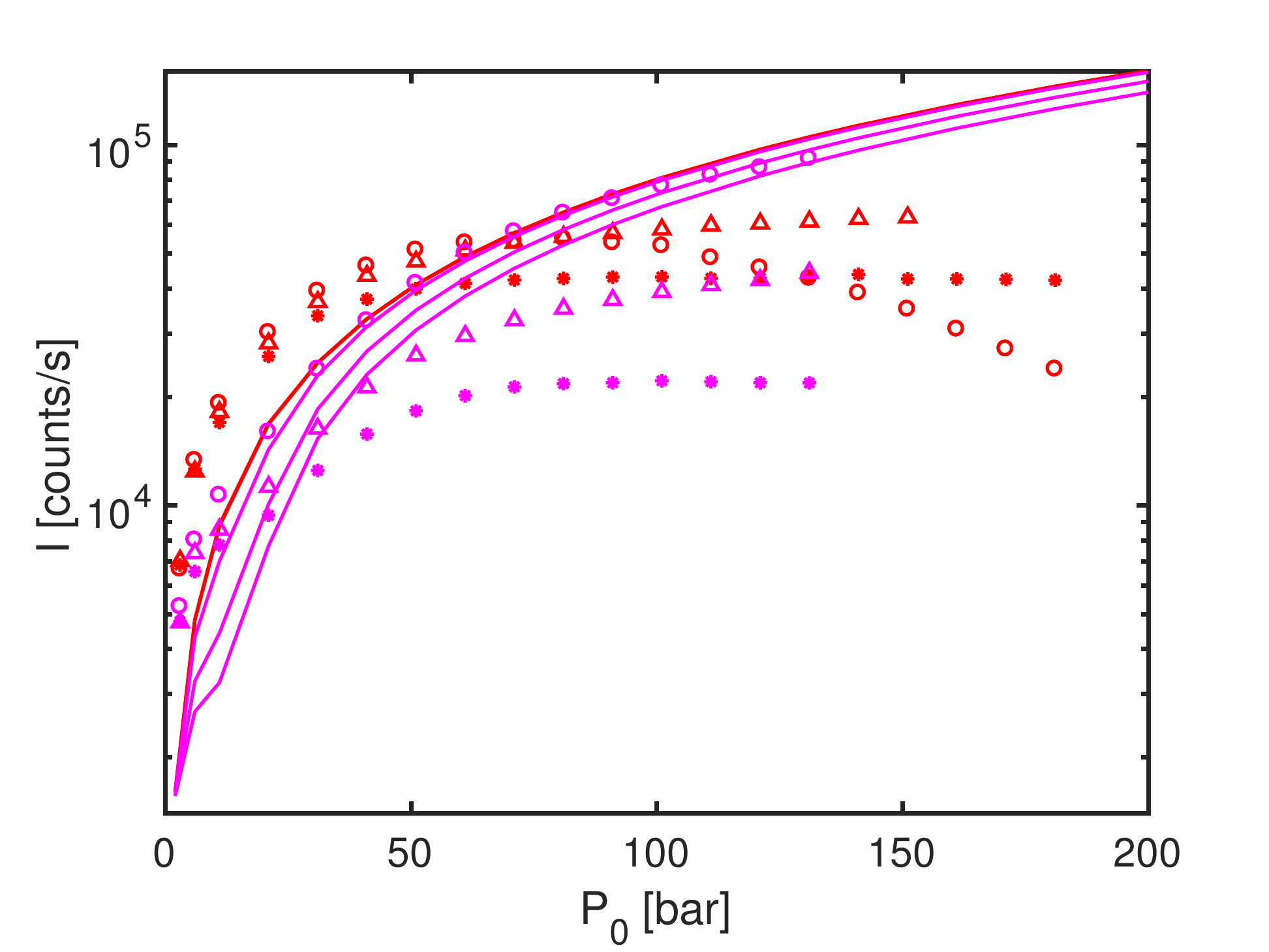}
	\caption{Plot of measured and predicted intensities for a cold source (125 K) and the Beam Dynamics skimmers: 120$\;\upmu\mathrm{m}$ (pink) and 390$\;\upmu\mathrm{m}$ (red). The intensities are computed using eq. (\ref{eq:scattering_contribution}) and assuming that the expansion stops at the skimmer with $S_i=S_\bot$. The intensities are plotted for three values of $x_\mathrm{S}$: 5.3 mm (circles), 11.3 mm (upwards arrows) and 17.3 mm (asterisks). Note how for $P_0>40$ bar and 390$\;\upmu\mathrm{m}$ skimmer (red), in the case of $x_\mathrm{S}=$ 5.3 mm, skimmer effects are clearly present and the centre line intensity is significantly lower than for the other two $x_\mathrm{S}$ positions. All measurements were taken with  $r_\mathrm{D}=25\upmu\mathrm{m}$ (see Appendix D and Fig. \ref{fig:exp_setup})\label{fig:cold_both}}
\end{figure}

We present the measured intensities for a beam with a source temperature of $125\pm2$ K and we compare them with the predictions from eq. (\ref{eq:scattering_contribution}). We obtain $n_\mathrm{B_E}\rightarrow n_\mathrm{B_E}(P_0)$ as in eq. (\ref{eq:densities}): $m_E=5\cdot 10^{-4} \frac{Pa}{bar}$, $n_E=48\cdot 10^{-4} Pa$. In the case of cold sources, if one chooses to determine the quitting surface position by the ratio of temperatures $T_\bot/T_\parallel\leq 0.1$, the quitting surface is placed after the skimmer already at quite low pressures. Thus, computing the eq. (\ref{eq:scattering_contribution}) for the case of $S_i=S_\parallel$ and the expansion stopping before the skimmer is only valid for a few measurement points. Therefore, we only present the results for the case of the expansion stopping at the skimmer and $S_i=S_\bot$. In general, the prediction power of the model decreases for a cold source (see Fig. \ref{fig:cold_both}).

\subsection{Micro skimmers}
\begin{figure}[htb]
	\centering
		\includegraphics[width=1\linewidth]{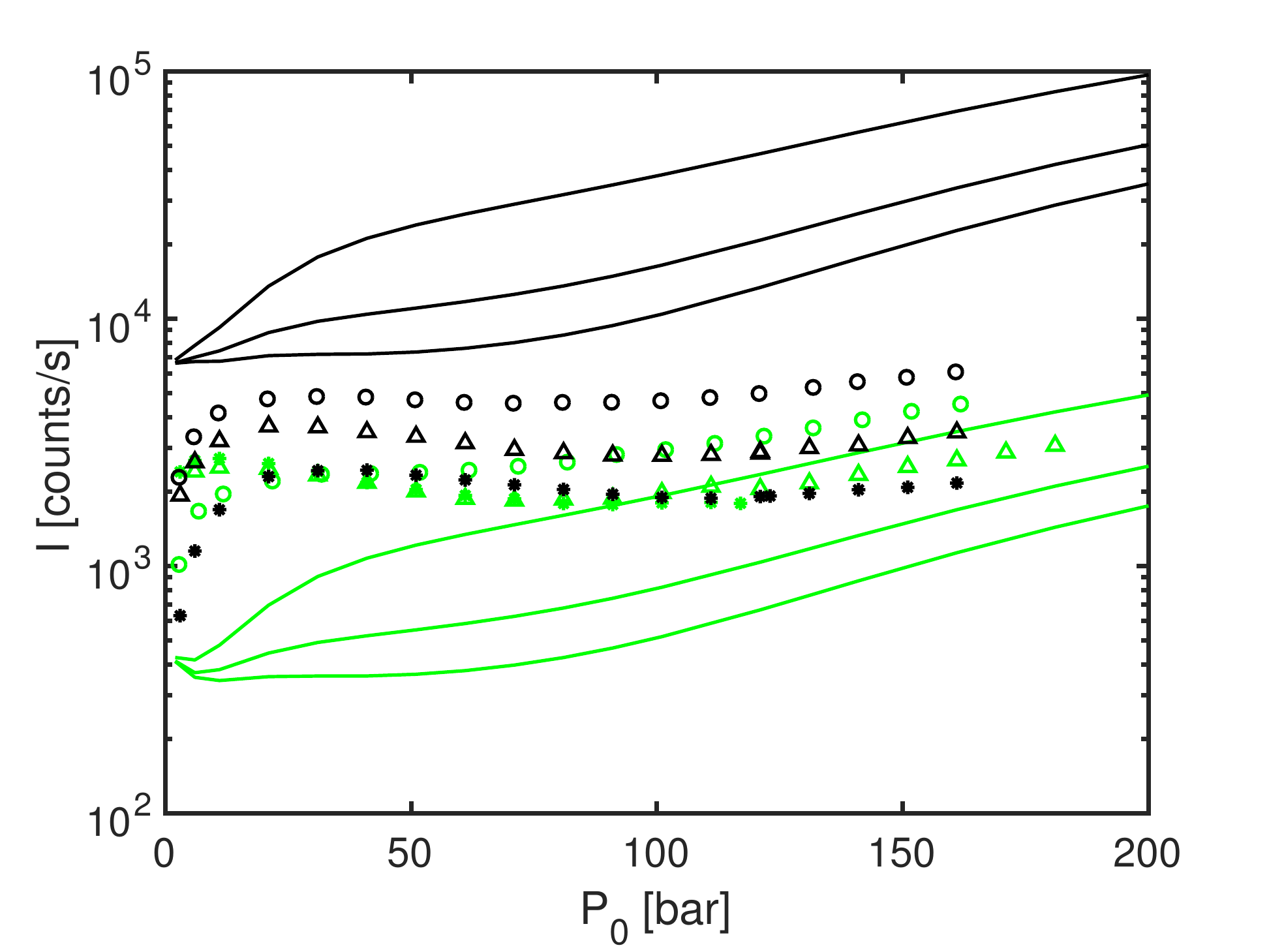}
	\caption{Plot of measured and predicted intensities for a warm source and the glass skimmers: 18$\upmu\mathrm{m}$ (black) and 4$\upmu\mathrm{m}$  (green). The intensities are computed using eq. (\ref{eq:scattering_contribution}) and assuming that the expansion stops at the skimmer with $S_i=S_\bot$.  \label{fig:microskimmer_warm}}
\end{figure}

\begin{figure}[htb]
	\centering
		\includegraphics[width=1\linewidth]{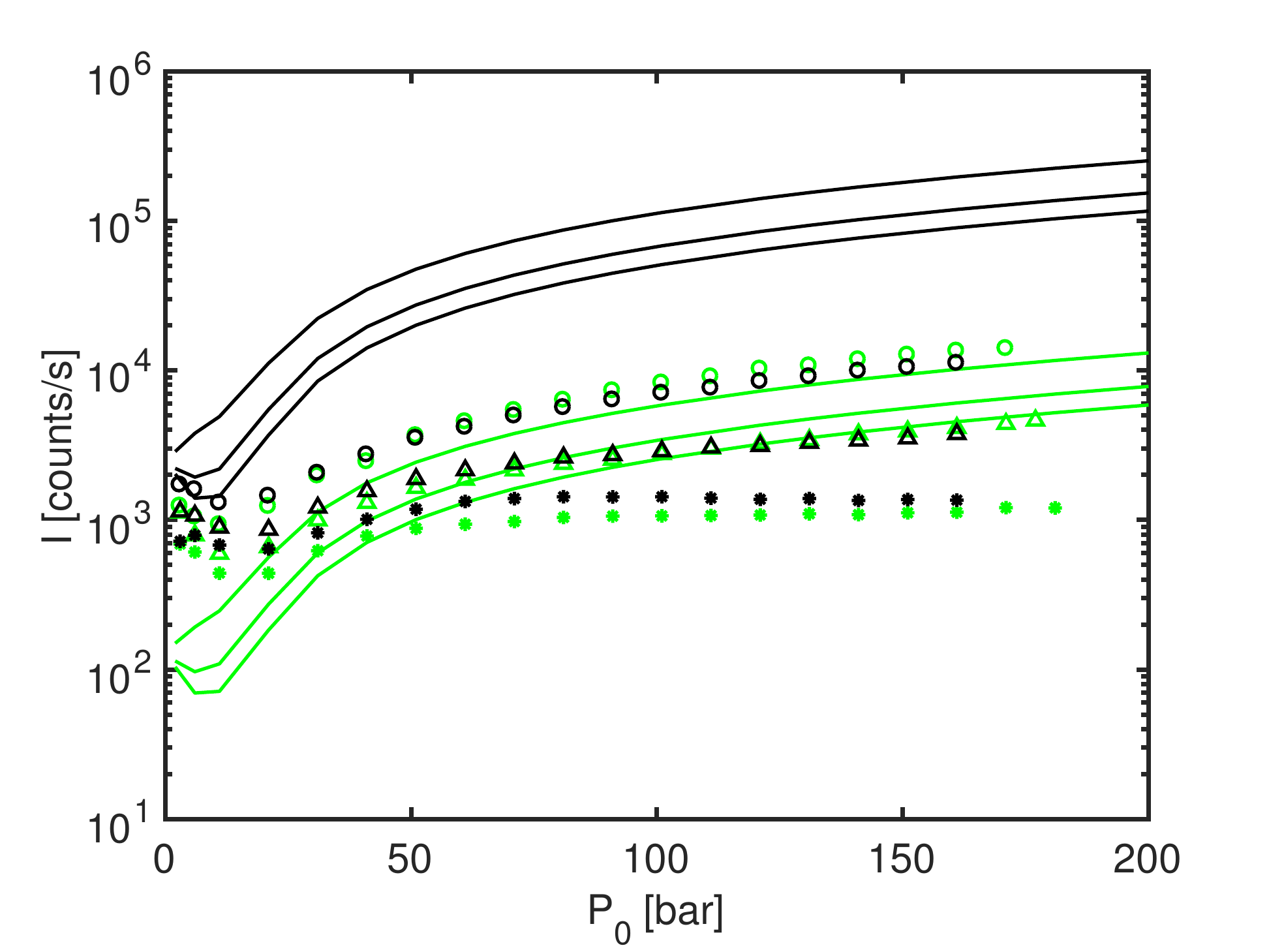}
	\caption{Plot of measured and predicted intensities for a cold source and the glass skimmers: 18$\upmu\mathrm{m}$ (black) and 4$\upmu\mathrm{m}$  (green). The intensities are computed  assuming that the expansion stops at the skimmer with $S_i=S_\bot$. \label{fig:microskimmer_cold}}
\end{figure}

The centre line intensity plots for micro skimmers show marked dips in the intensity, especially for the cold source cases. Centre line intensity dips are also observed at higher pressures for a warm source (see Figs. \ref{fig:microskimmer_warm} and \ref{fig:microskimmer_cold}). The model predicts the dips for a cold source, but in both cases fails to fit the experimental data well. The centre line intensity measured for both skimmers is in the same range, while the model predicts a more pronounced difference between the $18\;\upmu\mathrm{m}$ skimmer and the $4\;\upmu\mathrm{m}$ skimmer.

\subsection{The Kurt skimmer}
\begin{figure}[htb]
	\centering
		\includegraphics[width=1\linewidth]{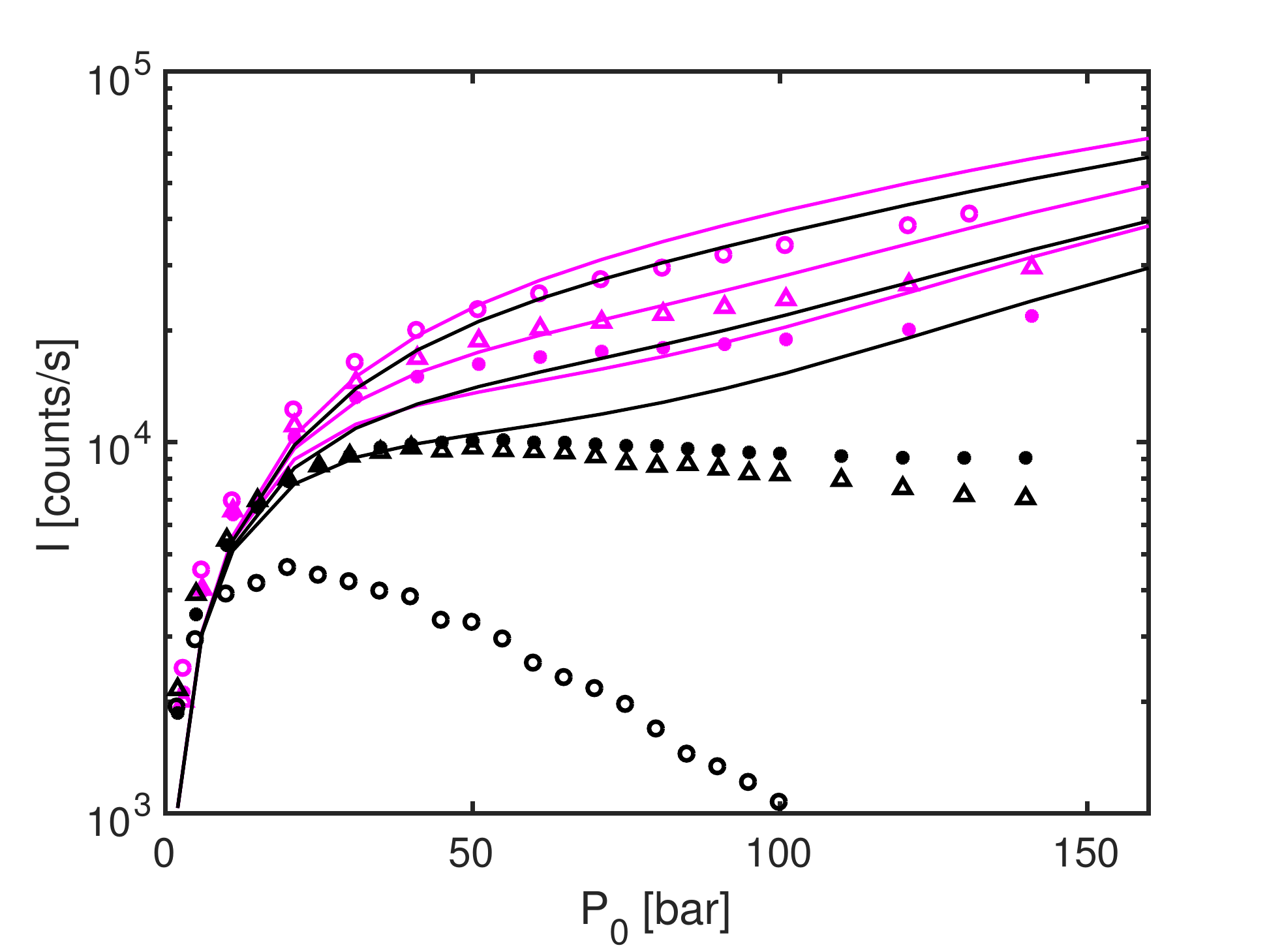}
	\caption{Plot of measured and computed intensities for the $100\;\upmu\mathrm{m}$ Kurt skimmer (black) and a 120 $\upmu\mathrm{m}$ (pink) Beam dynamics skimmer for a warm source. The intensities are computed  assuming that the expansion stops at the skimmer with $S_i=S_\bot$. Note how strong discrepancies are not observed except for the case of the $100\;\upmu\mathrm{m}$ Kurt skimmer. Two discrepancy modes can be observed, a very significant one for  $x_\mathrm{S}$= 5.3 mm and a less significant one for the rest of nozzle-skimmer distances. \label{fig:Kurt_effect}}
\end{figure}

To experimentally determine the importance of $\mathrm{Kn^*}$-driven skimmer effects we use a skimmer designed in such a way that such effects are expected to clearly dominate over the centre line intensity trends. This is the case of the Kurt skimmer (see Sec. \ref{sec:expmethod}), which due to its inverted-cone walls concentrates the reflecting particles along the beam center line, leading to a low $\mathrm{Kn^*}$ (see eq. (\ref{eq:mod_knudsen})). Comparing the Kurt skimmer intensities with the Beam dynamics skimmers, one sees that skimmer effects are not clearly observed until about 40 bar, for nozzle-skimmer distances corresponding to $x_\mathrm{S}>11.3$ mm (see Fig \ref{fig:Kurt_effect}). This means that the discrepancies at lower pressures between eq. (\ref{eq:scattering_contribution}) and the micro-skimmer measurements cannot be explained by skimmer interactions only. In fact, the modified Knudsen number in the case of micro-skimmers at 40 bar is expected to be larger than in the case of the Kurt skimmer due to the $1/r_\mathrm{S}$ dependency (see eq. (\ref{eq:mod_knudsen})). 

Note how skimmer interference in the case of the Kurt skimmer is not significant until the nozzle-skimmer distance is set at $5.3$ mm,  (see Fig \ref{fig:Kurt_effect}). A similar effect is seen, for a cold source, in the case of the 390 $\upmu\mathrm{m}$ Beam Dynamics skimmer, where for $x_\mathrm{S}=5.3$ mm, skimmer interference becomes evident (see Fig. \ref{fig:cold_both}). The same effect is not clearly observed for the smaller, 120 $\upmu\mathrm{m}$ Beam Dynamics skimmer. This can be seen as an experimental confirmation of the importance of the modified Knudsen number, which predicts stronger skimmer effects for larger skimmers.

\subsection{Complete experimental data}
\begin{figure}[htb]
	\centering
		\includegraphics[width=1\linewidth]{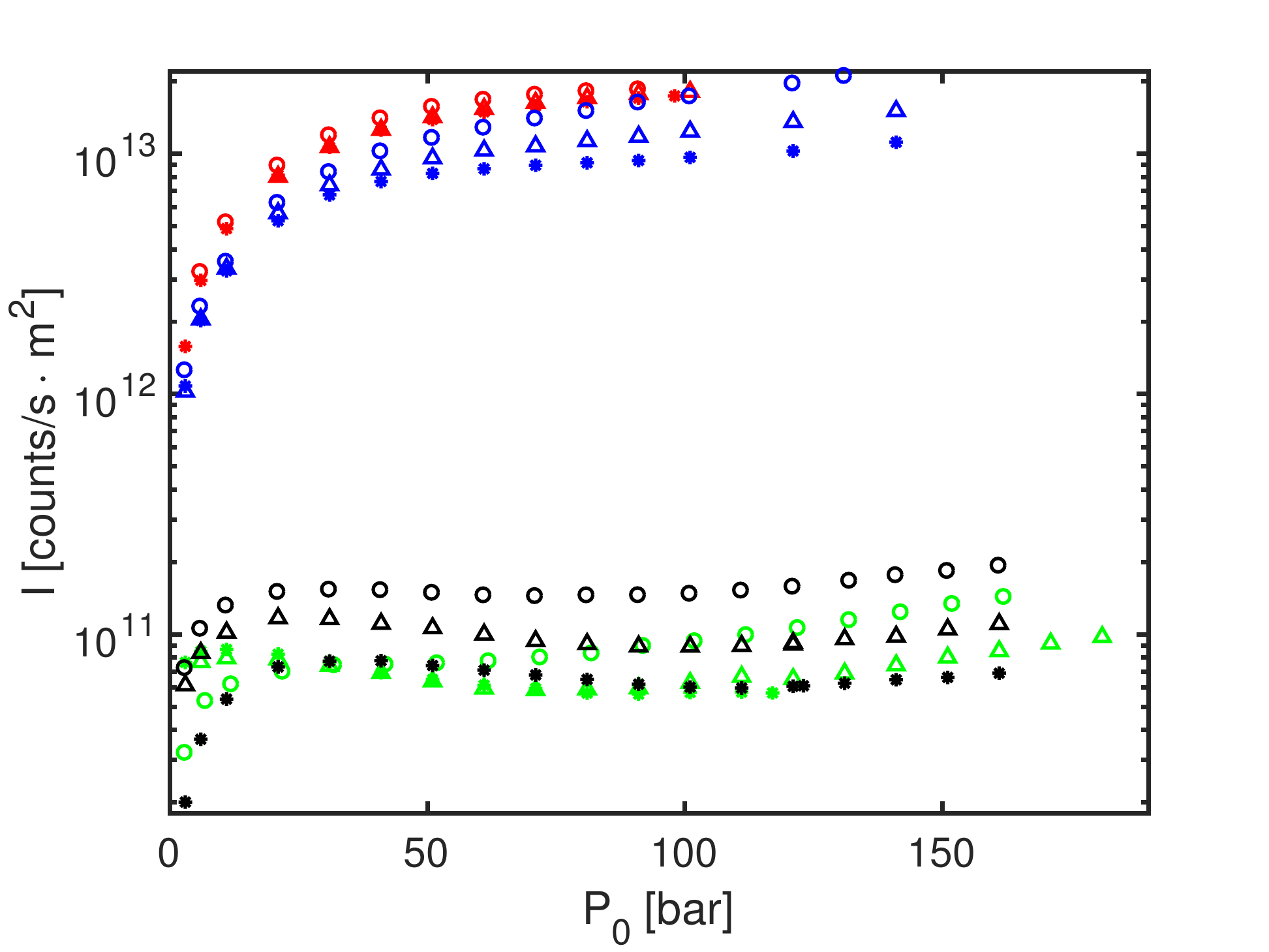}
	\caption{Measured centre line intensities per area in counts/$s\cdot m^2$ for a warm source, and for the following skimmer apertures: 120 $\upmu\mathrm{m}$ Beam Dynamics (blue), 390  $\upmu\mathrm{m}$ Beam Dynamics (red), 18 $\upmu\mathrm{m}$ glass skimmer (black), and $4\;\upmu\mathrm{m}$ glass skimmer (green). The circle, triangle, and asterisk markers correspond to the nozzle-skimmer distances, $x_\mathrm{S}$, of 5.3 mm, 11.3 mm, and 17.3 mm respectively.  \label{fig:all_warm}}
\end{figure}

\begin{figure}[htb]
	\centering
		\includegraphics[width=1\linewidth]{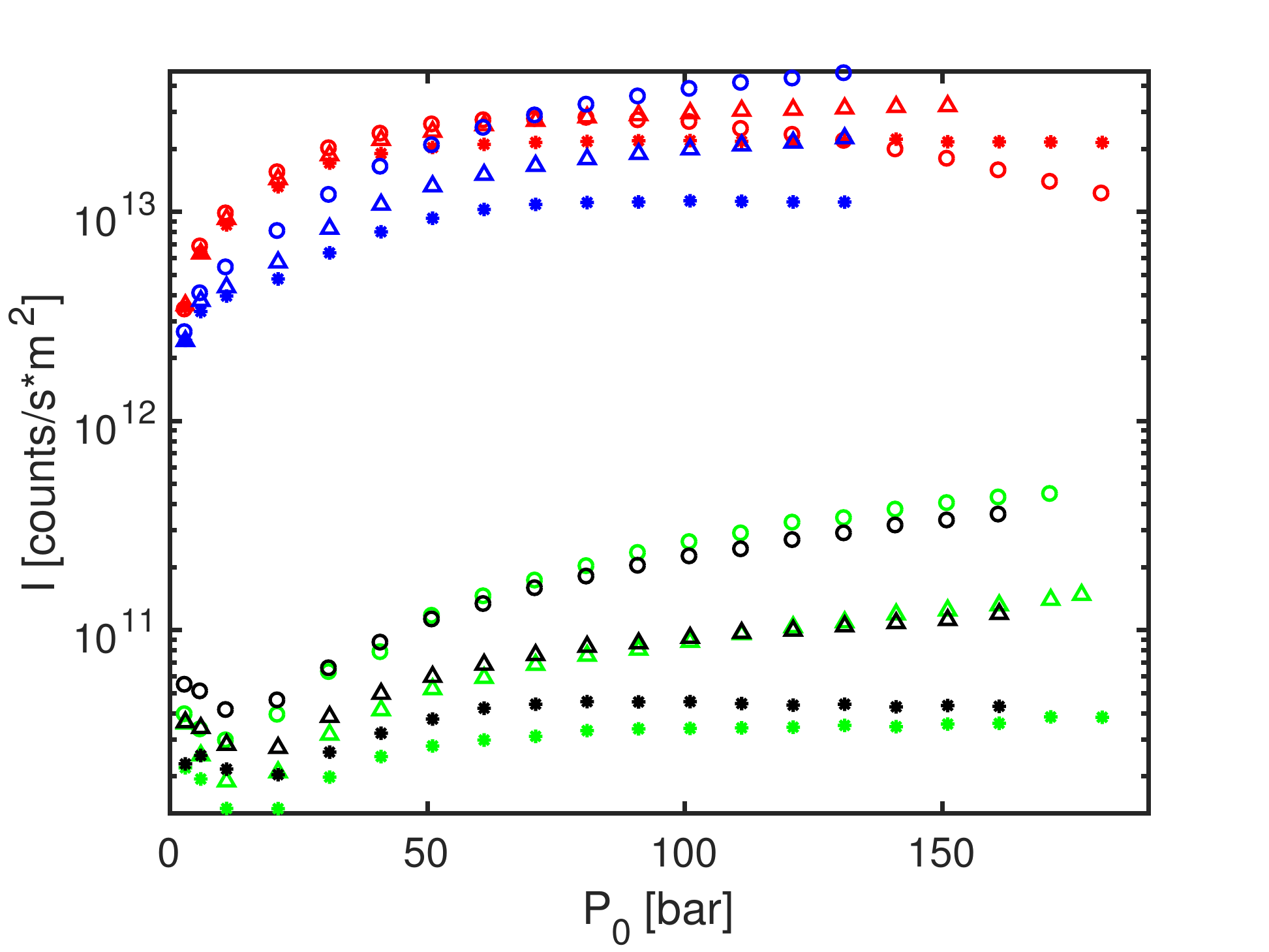}
	\caption{Measured centre line intensities per area in counts/$s\cdot m^2$ for a cold source, and for the following skimmer apertures: 120 $\upmu\mathrm{m}$ Beam Dynamics (blue), 390  $\upmu\mathrm{m}$ Beam Dynamics (red), 18 $\upmu\mathrm{m}$ glass skimmer (black), and $4\upmu\mathrm{m}$ glass skimmer (green). The round, triangle, and asterisk markers correspond to the nozzle-skimmer distances, $x_\mathrm{S}$, of 5.3 mm, 11.3 mm, and 17.3 mm respectively.  \label{fig:all_cold}}
\end{figure}

\begin{figure}[htb]
	\centering
		\includegraphics[width=1\linewidth]{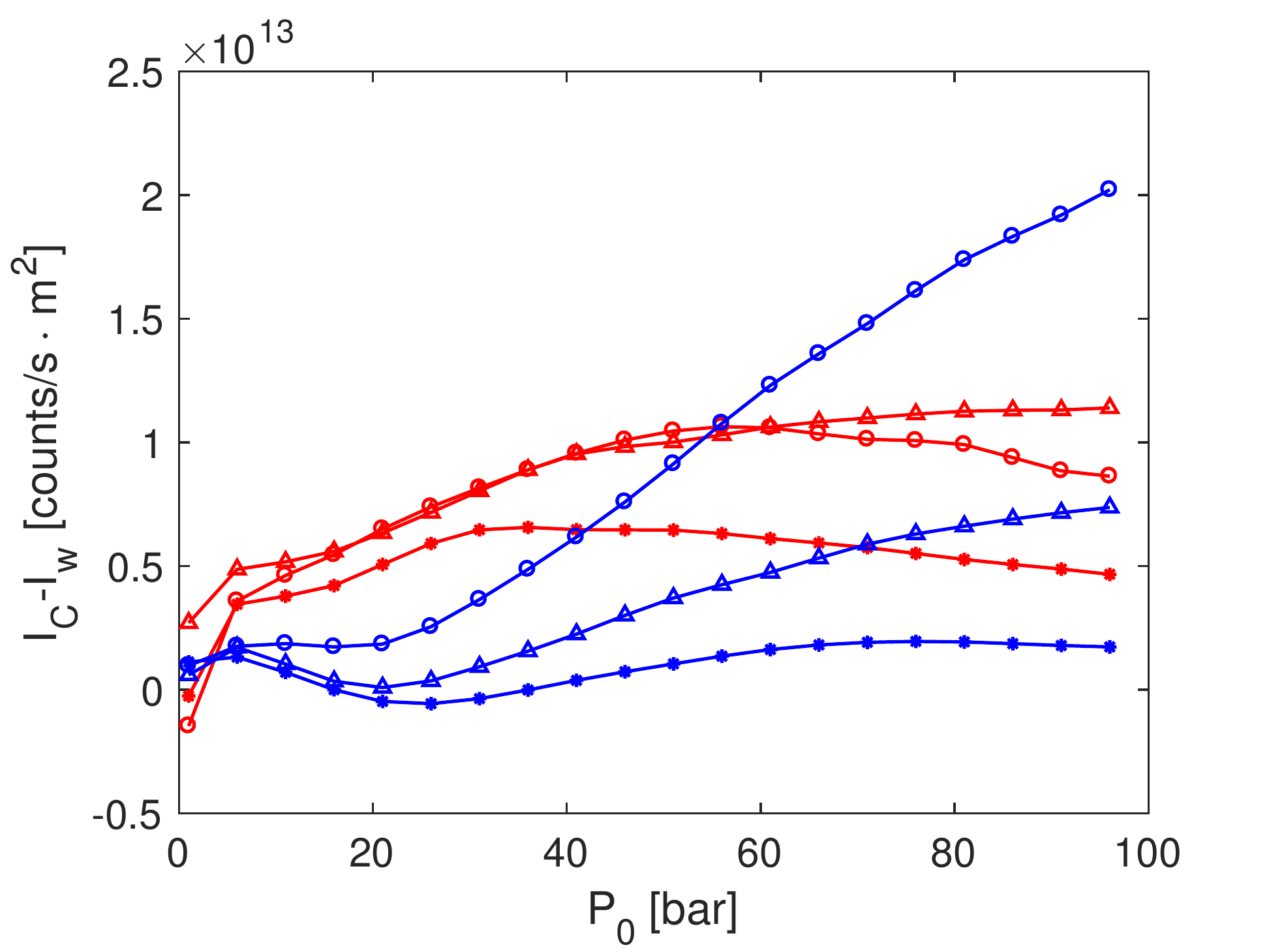}
	\caption{Measured differences between cold source and warm source beam intensities per area in counts/$s\cdot m^2$ for the following skimmer apertures: 120 $\upmu\mathrm{m}$ Beam Dynamics (blue), 390  $\upmu\mathrm{m}$ Beam Dynamics (red).The circle, triangle, and asterisk markers correspond to the nozzle-skimmer distances, $x_\mathrm{S}$, of 5.3 mm, 11.3 mm, and 17.3 mm respectively. The continuous line indicates that where experimental data was missing, data was extrapolated from the closest experimental points.  \label{fig:all_big}}
\end{figure}

\begin{figure}[htb]
	\centering
		\includegraphics[width=1\linewidth]{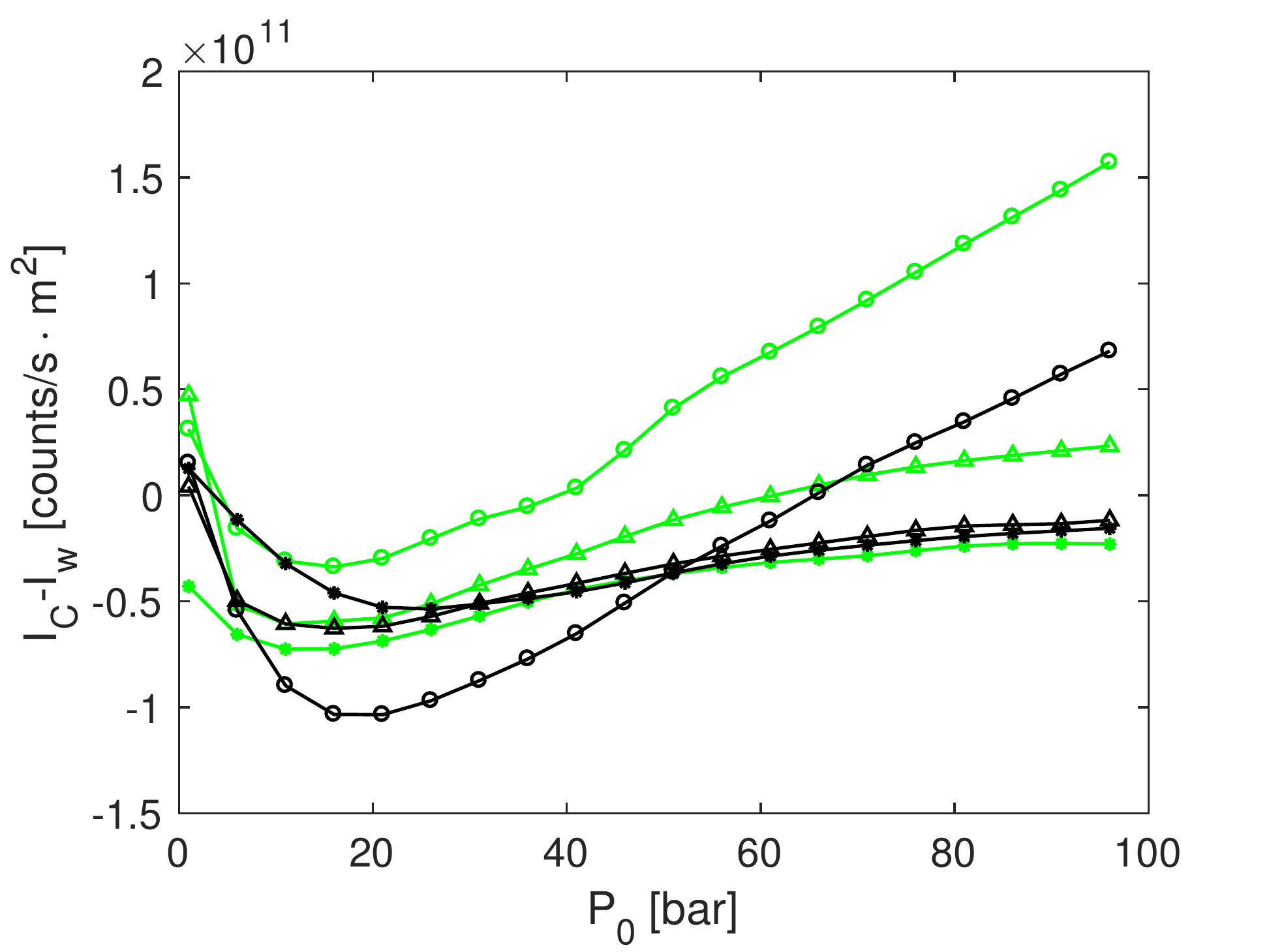}
	\caption{Measured differences between cold source and warm source beam intensities per area in counts/$s\cdot m^2$ for the following skimmer apertures: 18 $\upmu\mathrm{m}$ glass skimmer (black), and $4\upmu\mathrm{m}$ glass skimmer (green). The round, triangle, and asterisk markers correspond to the nozzle-skimmer distances, $x_\mathrm{S}$, of 5.3 mm, 11.3 mm, and 17.3 mm respectively.  The continuous line indicates that where experimental data was missing, data was extrapolated from the closest experimental points. \label{fig:all_small}}
\end{figure}

In this section, we plot the complete dataset of measurements carried out during this study, with the exception of measurements corresponding to the Kurt skimmer, that are plotted separately. In order to preserve the relevant intensity magnitude, and thus make comparisons easier the intensities plotted have been normalized to the radius of the aperture in front of the detector used to perform each measurement. Therefore, in this section, the intensities are given in $\mathrm{counts/s}\cdot \mathrm{m}^2$. The centre line intensity data for a warm source $T_0\approx 300$ K is shown in Fig. \ref{fig:all_warm}, and for a cold source  $T_0\approx 125$ K in Fig. \ref{fig:all_cold}. \textcolor{black}{Additionally, we plot the difference in centre line intensity per square meter between cold and warm sources for each experiment (Figs. \ref{fig:all_big} and \ref{fig:all_small}).}

\textcolor{black}{From Fig. \ref{fig:all_big}, one can observe that for large skimmers cold sources produce a higher centre line intensity than warm sources, especially for high source pressures. This is given by eq. (\ref{eq:intensity_nozzle}) and by the larger speed ratios obtained in cold beams. For the case of the 120$\upmu \mathrm{m}$ skimmer, this difference reduces the further away the skimmer is placed from the nozzle due to the evolution of $T_\bot$ along the beam axis.}

\textcolor{black}{For the case of micro skimmers,  cold sources are generally \emph{less} intense than warm sources, except for very large pressures. This is due to an intensity dip occurring for cold sources at low and medium pressures driven by the evolution of the beam's perpendicular speed (see Discussion). The smaller the collimating skimmer is the larger the influence of this dip on the measured centre line intensity. This is because larger skimmers collect particles with a larger perpendicular temperature range. }

\section{DISCUSSION}\label{sec:discussion}
The analytical model based on Sikora's ellipsoidal distribution approach ($S_i=S_\bot$, expansion stopped at the skimmer) predicts the centre line intensity of a helium beam generated by a source at ambient temperature with reasonable accuracy. However, the model has several  limitations, each of which will be discussed in detail in this section. 
\begin{enumerate}
\item \emph{Poor fit at high pressures}: for most skimmers, the model overshoots the measured intensities at high pressures ($P_0\gtrsim 50$ bar). This phenomenon is likely due to a combination of two effects: skimmer interference, and a continuing expansion of the beam after the skimmer. By observing the data, we can see that in the case of a warm source this overshoot does not significantly vary when two skimmers with the same design but different diameter are used (in this case, the Beam Dynamics skimmers). This points towards the idea that skimmer interference can't be the main cause of the overshoot, as the influence of the particles reflected from the skimmer is expected to strongly depend on the skimmer radius. However, in the case of a cold source, the overshoot is more significant for the $120$ $\upmu\mathrm{m}$ Beam Dynamics skimmer than its $390$ $\upmu\mathrm{m}$ equivalent. What is likely happening is that the helium beam continues to expand significantly after the skimmer following different dynamics than before it, due to the removal of particles by the skimmer edges. According to the simulations of the expansion performed in this study, this is particularly relevant for the case of a cold source, where the quitting surface is often predicted to be several centimetres after the skimmer. This renders Sikora's treatment of a beam that expands due to its non vanishing $T_\bot$ at the skimmer un-physical as it assumes no further collisions after the skimmer.

During the preparation of this paper, efforts were undertaken to adapt Sikora's model to a beam expanding after the skimmer using simple geometrical rules. This was motivated by the observations made by Doak et al, whom used micro-skimmers to perform focusing experiments  and observed a deviation between expected and measure focal spot size. They suggested that this may have been due to the supersonic expansion continuing after the beam has passed through the skimmer aperture \cite{PhysRevLett_4229}. This adaptation can be found in  Appendix A (Fig. \ref{fig:WARM_EXPAFTER}) but did not produce very promising results. A treatment using a DSMC simulation of the whole system is most likely a more accurate approach in order to predict intensities at large pressure values. This approach is also much more complex than the analytical models presented here.

Another possible explanation of these discrepancies would be the non-physical nature of a ``hard" quitting surface. Replacing it with a ``soft" treatment may yield interesting results. The centre line intensity would be calculated then by integrating over a series of infinitesimally spaced successive quitting surfaces.

The higher overshoot at $P_0\gtrsim 50$ bar for the smaller Beam Dynamics skimmer in the case of a cold source occurs in all cases except one: $x_\mathrm{S}=5.3$ mm (see Fig. \ref{fig:cold_both}).  In order to understand this peculiarity, one must re-visit the modified Knudsen number. The case of  $x_\mathrm{S}=5.3$ mm for a cold source and $r_\mathrm{S}=390\;\upmu\mathrm{m}$, is the case expected to have the lowest modified Knudsen number (largest $r_\mathrm{S}$ and number density at the skimmer, see eq.  (\ref{eq:mod_knudsen})). Therefore, it is likely that this particular case is the only one showing skimmer interference governed by the interaction with reflected particles.
\item \emph{Low predictability of micro-skimmer intensities}: on the one hand, skimmer interference and skimmer clogging are known to be determined by the modified Knudsen number $\mathrm{Kn^*}$, which strongly depends on the skimmer diameter (eq. (\ref{eq:mod_knudsen})). Micro-skimmers, are thus expected to show less interference than their larger counter-parts under the same conditions. This effect is clearly seen in Fig. \ref{fig:cold_both}, where skimmer effects are present only for the larger $390$ $\upmu\mathrm{m}$ skimmer. 

On the other hand, smaller skimmers sometimes have very thin and long geometries, causing a possible increase of pressure along the skimmer channel. This effect is likely what causes the bad fit between the model predictions and the observed micro-skimmer centre line intensities.

Notwithstanding, it is important to note that  Sikora's ellipsoidal quitting surface model is able to predict the \emph{general trends} of micro skimmer intensities. This includes the centre line intensity dip at low pressure for small skimmers. This dip is driven by the behaviour of the perpendicular speed ratio at low pressures, that is predicted by the simulation of the supersonic expansion to decrease first and increase later (see Fig. \ref{fig:Sbot_cold}). 

However, the experimental observability of this dip is actually determined by the radius of the skimmer and the distance between the nozzle and the skimmer. If $\frac{r_\mathrm{S}}{x_\mathrm{S}}S_\bot$ is small enough ($\lesssim 0.8$), then the term $\left[-S_\bot^2\left(\frac{r_\mathrm{S}(R_\mathrm{F}+a)}{R_\mathrm{F}(R_\mathrm{F}-x_\mathrm{S}+a)}\right)^2\right]$ in eq. (\ref{eq:I_sikora}) is small too. This makes the exponential term in eq. (\ref{eq:I_sikora}) dominate, and the effect of the dip in $S_\bot$ can be clearly observed in the beam centre line intensity. This explains why this dip is only experimentally observed for the case of micro-skimmers.

\begin{figure}[htb]
	\centering
		\includegraphics[width=0.7\linewidth]{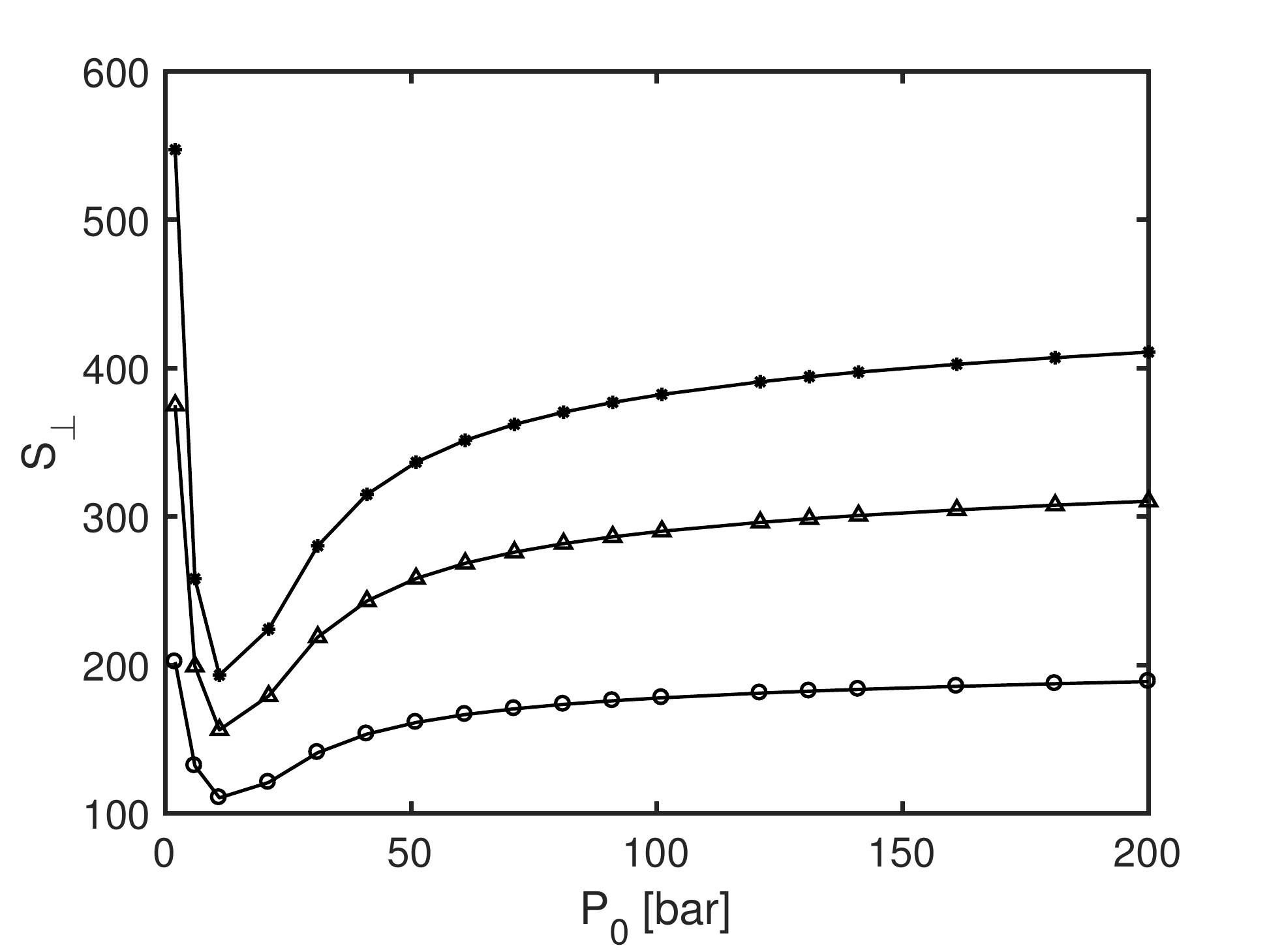}
	\caption{Predicted value of $S_\bot$ for a cold source (125 K) according to the numerical calculation of the supersonic expansion presented in Sec. \ref{sec:supersonic_exp}. The round, triangle, and asterisk markers correspond to the nozzle-skimmer distances, $x_\mathrm{S}$, of 5.3 mm, 11.3 mm, and 17.3 mm respectively.\label{fig:Sbot_cold}}
\end{figure}

This good trend replication is particularly relevant for purposes of optimization, where the value of interest is not so much the centre line intensity but the combination of parameters maximizing it.

\item \emph{Weak dependence on the nozzle-skimmer distance of the $S_i=S_\parallel$ variant}: only when the expansion is allowed to stop at the skimmer and the perpendicular speed ratio is used, does the predicted centre line intensity significantly depend on the nozzle-skimmer distance, $x_\mathrm{S}$. This is expected, as in this case the thermal spread of the beam is caused by the value of the perpendicular temperature at the skimmer $T_\bot$, and this value varies strongly with $x_\mathrm{S}$. \textcolor{black}{Despite $S_\parallel<<S_\bot$ causing a stronger exponential contribution in eq.  (\ref{eq:I_sikora}), the variation on $S_\bot$ with the skimmer radius is much stronger than the fraction term in the exponential, making the $S_i=S_\parallel$ variant actually less dependent on $x_\mathrm{S}$ (as $S_\parallel$ remains constant).}
\end{enumerate}
\section{CONCLUSION}
We present a dataset of centre line intensity measurements for a supersonic helium beam and compare it to various intensity models. We show that these models replicate the experimental data well for skimmers with diameters $120$ and $390~\upmu\mathrm{m}$. Particularly, we show that  Sikora's ellipsoidal distribution approach, assuming a quitting surface placed at the skimmer position, with the expansion dominated by the supersonic expansion perpendicular temperature $T_\bot$ fits the experimental data best.

We present a ray tracing simulation approach, used to numerically replicate the introduced centre line intensity models. We show that the ray tracing approach and analytical models (Sikora's and Bossel's) follow very similar dependencies with the different geometrical variables of the experiment.

In the presented dataset, we observe Knudsen number dependent skimmer interference for a $390$ $\upmu\mathrm{m}$ skimmer, and a specially designed $100\;\upmu\mathrm{m}$ skimmer placed 5.3 mm away from a cold source. We postulate that the rest of the discrepancies between the experimental data and the model may be due to either backscattering interferences at quasi-molecular flow regimes, or a continuation of the supersonic expansion after the beam has passed through the skimmer. Another explanation may be that the assumption of the quitting surface stopping abruptly at a given distance is is too simple to adequately describe the physics in this regime.

\begin{acknowledgments}
The work presented here was sponsored by the European Union: Theme NMP.2012.1.4-3 Grant no. 309672, project NEMI (Neutral Microscopy). We thank Yair Segev from the the Weizmann Institute of Science (Israel) for his very useful and detailed comments, especially for his comments regarding theoretical background. We thank Kurt Ansperger Design, Entwicklung und Bau von Prototypen, Moserhofgasse 24 C / 1, 8010 Graz,
Austria and Department of Physics/Experimental Physics, Fine-mechanical Laboratory and Workshop,
Karl-Franzens-University Graz, Universitatsplatz 5, 8010 Graz, Austria for the design and production of the Kurt skimmer. We thank Jon Roozenbeek for his useful edits.
\end{acknowledgments}

\section*{APPENDIX A: adaptation to an expansion after the skimmer}\label{sec:adapt}
\begin{figure}[htb]
	\centering
		\includegraphics[width=1\linewidth]{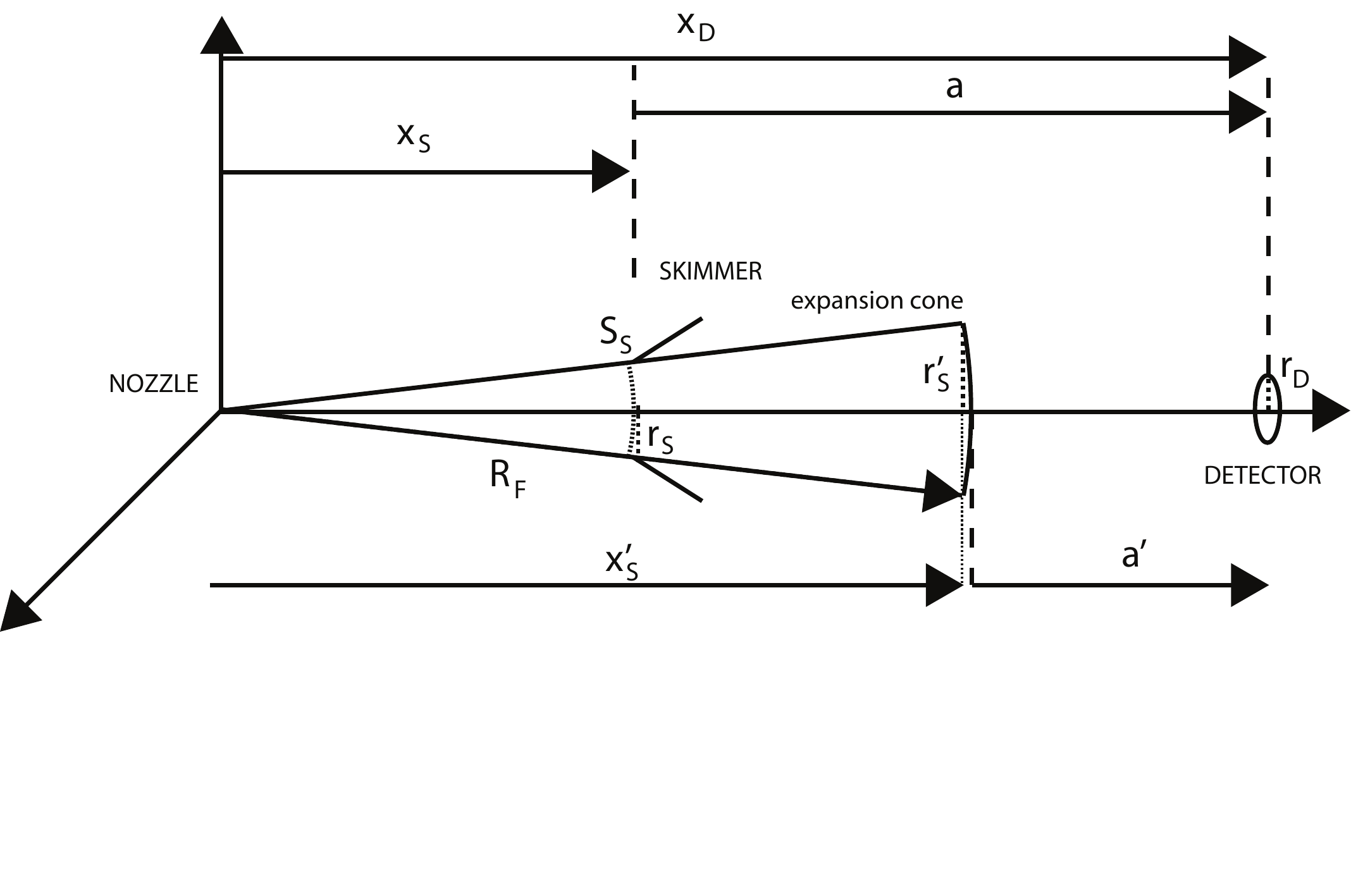}
	\caption{ Diagram of the supersonic expansion for the case of a radius of the quitting surface radius higher that the distance between the nozzle and the skimmer. The quitting surface is assumed to expand unaffected by the skimmer aperture, except by collimation. $R_\mathrm{F}$ is the radius of the quitting surface, $y$ is the distance between the axis of symmetry and the projection of the maximum-angle ray on the quitting surface, $r_\mathrm{S}$ is the skimmer radius and $r_\mathrm{D}$ is the radius of the detector. $a$ is the distance between the skimmer and the detector, $d$ is the distance from the skimmer to the point where the maximum-angle ray crosses the symmetry axis. $x_\mathrm{D}$ is the distance between the nozzle and the detector and $x_\mathrm{S}$ is the distance between the nozzle and the skimmer.  \label{fig:Q_S_expanding_afterskimmer}}
\end{figure}

An untreated case in literature is when collisional expansion continues after the skimmer. A way to approach this problem is to assume that the expansion is unaffected by this interaction and simply project the quitting surface further ahead until its predicted radius $R_\mathrm{F}$ (see Fig. \ref{fig:Q_S_expanding_afterskimmer}).

The centre line intensity must be calculated using eq. (\ref{eq:scattering_contribution}), with $a\rightarrow a'$, $r_\mathrm{S}\rightarrow r_\mathrm{S}'$, $x_\mathrm{S}\rightarrow x_\mathrm{S}'$:
\begin{equation}
a'=a-\left(R_\mathrm{F}\cos(\arctan\frac{r_\mathrm{S}}{x_\mathrm{S}})-x_\mathrm{S}\right)
\end{equation}
\begin{equation}
r_\mathrm{S}'=R_\mathrm{F}\sin\left(\arctan\frac{r_\mathrm{S}}{x_\mathrm{S}}\right)
\end{equation}
\begin{equation}
x_\mathrm{S}'=R_\mathrm{F}\cos\left(\arctan\frac{r_\mathrm{S}}{x_\mathrm{S}}\right)
\end{equation}

\begin{figure}[htb]
	\centering
		\includegraphics[width=1\linewidth]{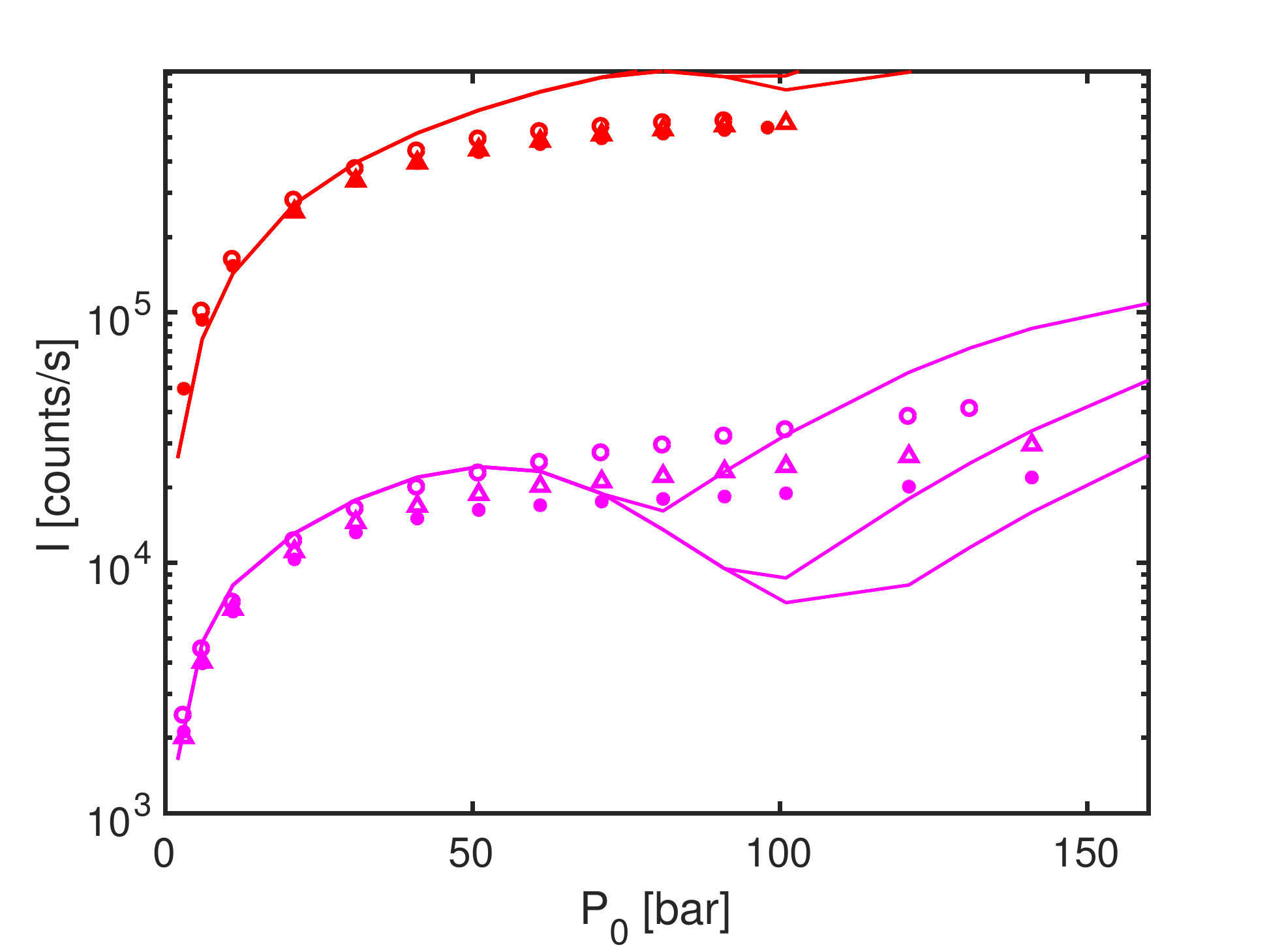}
	\caption{Plot of measured intensities for a warm source and the Beam Dynamics skimmers: (300 K), and 120$\upmu\mathrm{m}$ (pink) and 390$\upmu\mathrm{m}$ Beam Dynamics skimmers (red). The measured intensities are compared to eq. (\ref{eq:scattering_contribution}), with the expansion stopped after the skimmer and $S_i=S_\parallel$.   \label{fig:WARM_EXPAFTER}}
\end{figure}
\section*{APPENDIX B: derivation of the QS model}\label{sec:App}
The contribution to the number density by a differential of the quitting surface $dS$ placed at a point $P$ to the point $P'$ is \cite{bossel1974skimming}:
\begin{equation}\label{eq:dif_N}
dN(x_\mathrm{D},0,z_\mathrm{D})=n(R_\mathrm{F},\delta,\eta)f_\mathrm{ell}(v,\theta)d^3v.
\end{equation}
In this equation, $n(R_\mathrm{F},\delta,\eta)\equiv n(R_\mathrm{F})g(\delta)$ is the number density at the quitting surface, that is allowed to depend on the angle $\delta$ to account for the fact that the nozzle is not actually point-like. $f_\mathrm{ell}(v,\theta)$ is the ellipsoidal Maxwellian distribution defined in eq. (\ref{eq:ellipsoidal_vel_distribution}). $v$ is the modulus of the speed vector and $\theta$ is the angle between the segment PP' and P (see Fig. \ref{fig:sketch_ellipsoidal}). Following the derivation from \cite{bossel1974skimming}, one obtains:
\begin{multline}\label{eq:number_density_ataradius}
N(\mathrm{P}')=\\ \frac{\tau n(R_\mathrm{F})}{2\pi a^2}\int_0^{r_\mathrm{S}}\int_0^{\pi} g(\delta)r\cos^3\beta\cdot \epsilon^3e^{-S_\parallel^2(1-\epsilon^2\cos^2\theta)}D(b)dr d\alpha,
\end{multline}
where $S_\parallel=U/c_\parallel$ is the parallel speed ratio, $\epsilon=\left((\tau\sin^2\theta+\cos^2\theta\right)^{-1/2}$, $\tau=\frac{T_\parallel}{T_\bot}$.
The function $D(b)$ is defined as follows:
\begin{equation}
D(b)\equiv \frac{2}{\sqrt{\pi}}be^{-b^2}+\left(2b^2+1\right)\left[1+erf(b)\right],\qquad b\equiv S_\parallel\epsilon\cos\theta
\end{equation}
The angle $\beta$ is shown in Fig. \ref{fig:sketch_ellipsoidal}. $N(P')$ corresponds to the number density at a radial position from the axis of symmetry, to obtain the number density at a circular detector we must integrate over the arriving differential volume:
\begin{equation}
N_{\mathrm{total}}=\Delta x\int_\mathrm{S} N(\mathrm{P}')d\mathrm{S}= 2\pi\Delta x\int_0^{r_\mathrm{D}} N(x_\mathrm{D},\rho) \rho d\rho.
\end{equation}
Imposing that the proportion of intensities must correspond to the proportion of number densities, we can obtain the expression for the centre line intensity arriving at a circular detector:
\begin{equation}
\frac{I_{\mathrm{D}}}{I_0}=\frac{N_\mathrm{total}}{2\pi\int_{R_\mathrm{F}-\Delta x}^{R_\mathrm{F}}\int_0^{\frac{\pi}{2}}n(r)r^2 g(\delta)\sin\delta d\delta dr}.
\end{equation}
We obtain:
\begin{multline}\label{eq:intensity_zoneplate_second_norm}
I_{\mathrm{D}}=\frac{\tau I_0}{2\pi a^2 R_\mathrm{F}^2{L}}\int_0^{r_\mathrm{D}}\int_0^{r_\mathrm{S}}\int_0^{\pi} g(\delta)r\cdot\rho\cos^3\beta\cdot \epsilon^3 \\ e^{-S_\parallel^2(1-\epsilon^2\cos^2\theta)}D(b)d\rho dr d\alpha.
\end{multline}
Where $I_0$ is defined in eq. (\ref{eq:intensity_nozzle}). L corresponds to the integration of $g(\delta)$ along the half sphere (all the intensity emitted by the source is set to be contained in $g(\delta)$).
\begin{equation}\label{eq:L_definition}
L\equiv \int_0^{\frac{\pi}{2}} g(\delta)\sin\delta d\delta.
\end{equation}.

\section*{APPENDIX C: Equations for the ray tracing code\label{ap:Monte Carlo}}
Using trigonometry, it is possible to determine exactly the maximum possible $\delta_m$ within a source-skimmer-detector geometry (see Fig. \ref{fig:LIMIT_ANGLE}).
\begin{equation}
\delta_m=\arcsin\frac{y}{R_\mathrm{F}}.
\end{equation}
Now, we use the Pythagorean theorem to obtain $y$, the height of the triangle containing the angle $\delta_m$, $x$ is the basis of the triangle as shown in Fig. \ref{fig:LIMIT_ANGLE}.
\begin{equation}\label{eq:pythagoras}
\frac{y}{d+(x_\mathrm{S}-x)}=\frac{r_\mathrm{S}}{d},\qquad x=\sqrt{R_\mathrm{F}^2-y^2}.
\end{equation}
Expanding eqs. (\ref{eq:pythagoras}) we obtain the following quadratic equation:
\begin{equation}
\left(\frac{yd}{r_\mathrm{S}}-d-x_\mathrm{S}\right)^2=R_\mathrm{F}^2-y^2,
\end{equation}
expanding in powers of $y$:
\begin{equation}
y^2\left((\frac{d}{r_\mathrm{S}})^2+1\right)+y\left(-2\frac{d}{r_\mathrm{S}}(d+x_\mathrm{S}) \right)+\left( d+x_\mathrm{S}\right)^2-R_\mathrm{F}^2=0
\end{equation}
Which can be solved using the quadratic formula:
\begin{multline}\label{eq:quadratic_geometric}
y=\frac{2\frac{d(d+x_\mathrm{S})}{r_\mathrm{S}}\pm\sqrt{4\frac{d^2}{r_\mathrm{S}^2}R_\mathrm{F}^2-4(d+x_\mathrm{S})^2+4R_\mathrm{F}^2}}{2(\frac{d}{r_\mathrm{S}})^2+2}=\\ \frac{\frac{d}{r_\mathrm{S}}(d+x_\mathrm{S})\pm\sqrt{\frac{d^2}{r_\mathrm{S}^2}R_\mathrm{F}^2+R_\mathrm{F}^2-(d+x_\mathrm{S})^2}}{(\frac{d}{r_\mathrm{S}})^2+1}.
\end{multline}
The distance $d$ is also obtained using trigonometry (see Fig. \ref{fig:LIMIT_ANGLE}).
\begin{equation}
\frac{r_\mathrm{S}}{d}=\frac{r_\mathrm{D}}{a-d}\rightarrow d=\frac{a r_\mathrm{S}}{r_\mathrm{D}+r_\mathrm{S}}.
\end{equation}
To determine whether to take the positive or negative square root in eq. \ref{eq:quadratic_geometric}, we can take the case $x=R_\mathrm{F}$ (which corresponds to the case $R_\mathrm{F}\rightarrow\infty$). In this case, from trigonometry it is easy to see that $y=\frac{r_\mathrm{S}}{d}(d+x_\mathrm{s}-R_\mathrm{F})$. Thus, the geometrically-sound case corresponds to the negative square root.

\section{APPENDIX D: $r_\mathrm{D}$-$r_\mathrm{S}$ table}
\begingroup
\squeezetable
\begin{table}[H]
\centering
\caption{Table showing the values for the skimmer radius $r_\mathrm{S}$, and the radius of the pinhole placed in front of the detector $r_\mathrm{D}$, for the experiments presented in this paper.}
\label{table:radiuses}
\begin{tabular}{|l|l|l|l|}
\hline
Skimmer diameter      & $r_\mathrm{S}$        & $r_\mathrm{D}$ (warm) & $r_\mathrm{D}$ (cold) \\ \hline
4 $\upmu\mathrm{m}$   & 2 $\upmu\mathrm{m}$   & 100 $\upmu\mathrm{m}$ & 100 $\upmu\mathrm{m}$ \\ \hline
18 $\upmu\mathrm{m}$  & 9 $\upmu\mathrm{m}$   & 100 $\upmu\mathrm{m}$ & 100 $\upmu\mathrm{m}$ \\ \hline
100 $\upmu\mathrm{m}$ & 50 $\upmu\mathrm{m}$  & 25 $\upmu\mathrm{m}$  & not shown             \\ \hline
120 $\upmu\mathrm{m}$ & 60 $\upmu\mathrm{m}$  & 25 $\upmu\mathrm{m}$  & 25 $\upmu\mathrm{m}$  \\ \hline
390 $\upmu\mathrm{m}$ & 195 $\upmu\mathrm{m}$ & 100 $\upmu\mathrm{m}$  & 25 $\upmu\mathrm{m}$  \\ \hline
\end{tabular}
\end{table}
\mbox{} 
\endgroup
\section{APPENDIX E: glossary of symbols}
\begingroup
\squeezetable
\begin{longtable}{ll}
\textbf{Symbol}                      & Description                                                                                                                                                                                       \\[1pt] \hline
\endfirsthead
\endhead

&\\[1pt]
$\mathrm{Kn}^*$             & Modified Knudsen number                                                                                                                                                                           \\[1pt]
Kn                          & Knudsen number                                                                                                                                                                                    \\[1pt]
$S_{\parallel}$             & Parallel speed ratio                                                                                                                                                                              \\[1pt]
$\eta_\mathrm{P}$           & Power law parameter on the collision model                                                                                                                                                        \\[1pt]
$\lambda_0$                 & Mean free path of gas particles                                                                                                                                                                   \\[1pt]
$r_\mathrm{S}$              & Skimmer radius                                                                                                                                                                                    \\[1pt]
$n$                         & Number density of the gas at the skimmer                                                                                                                                                          \\[1pt]
$\sigma$                    & Cross section of gas particles                                                                                                                                                                    \\[1pt]
$\bar{v}$                   & \begin{tabular}[c]{@{}l@{}}Most probable velocity along \\ the radial direction\end{tabular}                                                                                                      \\
$T_\parallel$               & Parallel temperature of the expansion                                                                                                                                                             \\[1pt]
$T_\bot$                    & Perpendicular temperature of the expansion                                                                                                                                                        \\[1pt]
$v_\parallel$               & Parallel component of the velocity                                                                                                                                                                \\[1pt]
$v_\bot$                    & Perpendicular component of the velocity                                                                                                                                                           \\[1pt]
M                           & Mach number                                                                                                                                                                                       \\[1pt]
v                           & Average velocity of the gas                                                                                                                                                                       \\[1pt]
c                           & Local speed of sound                                                                                                                                                                              \\[1pt]
$I_0$                       & Total intensity stemming from the nozzle                                                                                                                                                          \\[1pt]
$T_0$                       & Stagnation temperature inside the nozzle                                                                                                                                                          \\[1pt]
$P_0$                       & Stagnation pressure inside the nozzle                                                                                                                                                             \\[1pt]
$k_\mathrm{B}$              & Boltzmann constant                                                                                                                                                                                \\[1pt]
$\gamma$                    & Ratio of heat capacities                                                                                                                                                                          \\[1pt]
$d_\mathrm{N}$              & Diameter of the nozzle                                                                                                                                                                            \\[1pt]
m                           & Mass of a gas particle                                                                                                                                                                            \\[1pt]
$\Omega (T_{\mathrm{eff}})$ & Collision integral in the Boltzmann equation                                                                                                                                                      \\[1pt]
$T_\mathrm{eff}$            & Effective average temperature of the gas                                                                                                                                                          \\[1pt]
$Q^2$                       & Viscosity cross-section                                                                                                                                                                           \\[1pt]
E                           & Collision energy in the centre of mass system                                                                                                                                                     \\[1pt]
$\hbar$                     & Reduced Planck constant                                                                                                                                                                           \\[1pt]
$\eta_l$                    & Phase shifts for orbital momentum $l$                                                                                                                                                             \\[1pt]
$f_\mathrm{ell}$            & Velocity distribution in the expansion                                                                                                                                                            \\[1pt]
$V_\mathrm{LJ}$             & Lennard-Jones potential                                                                                                                                                                           \\[1pt]
$r_\mathrm{LJ}$             & Distance between two interacting particles                                                                                                                                                        \\[1pt]
$r_\mathrm{m}$              & Distance where $V_\mathrm{LJ}$ is minimum                                                                                                                                                         \\[1pt]
$\epsilon$                  & Depth of the potential well in $V_\mathrm{LJ}$                                                                                                                                                    \\[1pt]
$R_\mathrm{F}$              & Radius of the quitting surface                                                                                                                                                                    \\[1pt]
$I_\mathrm{D}$              & Centre-line intensity (ellipsoidal model)                                                                                                                                                         \\[1pt]
$\tau$                      & $T_\parallel/T_\bot$                                                                                                                                                                              \\[1pt]
a                           & \begin{tabular}[c]{@{}l@{}}Distance between the skimmer \\ and the detector\end{tabular}                                                                                                          \\
$r_\mathrm{D}$              & Radius of detector opening                                                                                                                                                                        \\[1pt]
P                           & Point on the quitting surface                                                                                                                                                                     \\[1pt]
P'                          & Point on the detector                                                                                                                                                                             \\[1pt]
$\vec{r}$                   & Vector connecting P and P'                                                                                                                                                                        \\[1pt]
r                           & \begin{tabular}[c]{@{}l@{}}Distance from beam axis to where the \\ skimmer plane intersects $\vec{r}$\end{tabular}                                                                                \\
x,y,z                       & Cartesian coordinates                                                                                                                                                                             \\[1pt]
$\beta$                     & Angle between $\vec{r}$ and the xz plane                                                                                                                                                          \\[1pt]
$\theta$                    & Angle between P and $\vec{r}$                                                                                                                                                                     \\[1pt]
$\alpha$                    & \begin{tabular}[c]{@{}l@{}}Angle between r (note, not \textbackslash{}vec\{r\}) and \\ the xz plane\end{tabular}                                                                                  \\
$\rho$                      & Distance between P' and the detector centre                                                                                                                                                       \\[1pt]
$g(\delta)$                 & \begin{tabular}[c]{@{}l@{}}Angular dependency of the gas density \\ on the quitting surface\end{tabular}                                                                                          \\
$\mathrm{L}$                & Integral of $g(\delta)$ over the quitting surface                                                                                                                                                 \\[1pt]
$S_\mathrm{i}$              & Speed ratio term in Sikora's model                                                                                                                                                                \\[1pt]
$I$                         & \begin{tabular}[c]{@{}l@{}}Sikora's centre line intensity before\\ approximation\end{tabular}                                                                                                     \\
$\Phi$                      & Angle of rotation about the beam axis                                                                                                                                                             \\[1pt]
$T_{\parallel\infty}$       & Asymptotic value of the parallel temperature                                                                                                                                                      \\[1pt]
$\eta_\mathrm{D}$           & Efficiency of the detector in counts/part                                                                                                                                                         \\[1pt]
$I_1$                       & \begin{tabular}[c]{@{}l@{}}Intensity arriving at the detector assuming\\ no skimmer presence\end{tabular}                                                                                         \\
$x_\mathrm{S}$              & Distance between nozzle and skimmer                                                                                                                                                               \\[1pt]
$I_\mathrm{S}$              & \begin{tabular}[c]{@{}l@{}}Sikora's centre line intensity assuming \\ $ r_\mathrm{S}\ll x_\mathrm{S}, \; r_\mathrm{S}\ll a$,$a/ r_\mathrm{S}>>S_i$, $r_\mathrm{D}<<a$\end{tabular} \\[1pt]
$n_\mathrm{BE}$             & \begin{tabular}[c]{@{}l@{}}Background number density in the \\ expansion chamber\end{tabular}                                                                                                     \\
$n_\mathrm{BC}$             & \begin{tabular}[c]{@{}l@{}}Background number density in \\ subsequent chambers\end{tabular}                                                                                                       \\
$\delta_\mathrm{m}$         & Maximal angle on the quitting surface                                                                                                                                                             \\[1pt]
d                           & \begin{tabular}[c]{@{}l@{}}Distance from the skimmer to the point where \\ the maximum-angle ray crosses the beam axis\end{tabular}                                                               \\[1pt]
$\phi$                      & Azimuthal angle in spherical coordinates                                                                                                                                                          \\[1pt]
$\delta$                    & \begin{tabular}[c]{@{}l@{}}Polar angle in spherical coordinates \\ ($\delta=0$ lays over x)\end{tabular}                                                                                          \\[1pt]
$A_\mathrm{D}$              & Area of the detector                                                                                                                                                                              \\[1pt]
$A_\mathrm{S}$              & Area of the skimmer                                                                                                                                                                               \\[1pt]
$x_\mathrm{A}$              & \begin{tabular}[c]{@{}l@{}}Distance between the nozzle and the \\ noise-reducing aperture\end{tabular}                                                                                           
\end{longtable}
\endgroup
\bibliographystyle{aipnum4-1}

\begin{thebibliography}{43}%
\makeatletter
\providecommand \@ifxundefined [1]{%
 \@ifx{#1\undefined}
}%
\providecommand \@ifnum [1]{%
 \ifnum #1\expandafter \@firstoftwo
 \else \expandafter \@secondoftwo
 \fi
}%
\providecommand \@ifx [1]{%
 \ifx #1\expandafter \@firstoftwo
 \else \expandafter \@secondoftwo
 \fi
}%
\providecommand \natexlab [1]{#1}%
\providecommand \enquote  [1]{``#1''}%
\providecommand \bibnamefont  [1]{#1}%
\providecommand \bibfnamefont [1]{#1}%
\providecommand \citenamefont [1]{#1}%
\providecommand \href@noop [0]{\@secondoftwo}%
\providecommand \href [0]{\begingroup \@sanitize@url \@href}%
\providecommand \@href[1]{\@@startlink{#1}\@@href}%
\providecommand \@@href[1]{\endgroup#1\@@endlink}%
\providecommand \@sanitize@url [0]{\catcode `\\12\catcode `\$12\catcode
  `\&12\catcode `\#12\catcode `\^12\catcode `\_12\catcode `\%12\relax}%
\providecommand \@@startlink[1]{}%
\providecommand \@@endlink[0]{}%
\providecommand \url  [0]{\begingroup\@sanitize@url \@url }%
\providecommand \@url [1]{\endgroup\@href {#1}{\urlprefix }}%
\providecommand \urlprefix  [0]{URL }%
\providecommand \Eprint [0]{\href }%
\providecommand \doibase [0]{http://dx.doi.org/}%
\providecommand \selectlanguage [0]{\@gobble}%
\providecommand \bibinfo  [0]{\@secondoftwo}%
\providecommand \bibfield  [0]{\@secondoftwo}%
\providecommand \translation [1]{[#1]}%
\providecommand \BibitemOpen [0]{}%
\providecommand \bibitemStop [0]{}%
\providecommand \bibitemNoStop [0]{.\EOS\space}%
\providecommand \EOS [0]{\spacefactor3000\relax}%
\providecommand \BibitemShut  [1]{\csname bibitem#1\endcsname}%
\let\auto@bib@innerbib\@empty
\bibitem [{\citenamefont {Campargue}(1964)}]{Campargue1964}%
  \BibitemOpen
  \bibfield  {author} {\bibinfo {author} {\bibfnamefont {R.}~\bibnamefont
  {Campargue}},\ }\href {\doibase 10.1063/1.1718676} {\bibfield  {journal}
  {\bibinfo  {journal} {Rev. Sci. Instrum.}\ }\textbf {\bibinfo {volume}
  {35}},\ \bibinfo {pages} {111} (\bibinfo {year} {1964})}\BibitemShut
  {NoStop}%
\bibitem [{\citenamefont {Pauly}(2000{\natexlab{a}})}]{Pauly2000}%
  \BibitemOpen
  \bibfield  {author} {\bibinfo {author} {\bibfnamefont {H.}~\bibnamefont
  {Pauly}},\ }\href@noop {} {\emph {\bibinfo {title} {{Atom, Molecule, and
  Cluster Beams I}}}},\ \bibinfo {edition} {1st}\ ed.\ (\bibinfo  {publisher}
  {Springer-Verlag},\ \bibinfo {address} {Berlin},\ \bibinfo {year}
  {2000})\BibitemShut {NoStop}%
\bibitem [{\citenamefont {DePonte}, \citenamefont {Kevan},\ and\ \citenamefont
  {Patton}(2006)}]{DePonte2006}%
  \BibitemOpen
  \bibfield  {author} {\bibinfo {author} {\bibfnamefont {D.~P.}\ \bibnamefont
  {DePonte}}, \bibinfo {author} {\bibfnamefont {S.~D.}\ \bibnamefont {Kevan}},
  \ and\ \bibinfo {author} {\bibfnamefont {F.~S.}\ \bibnamefont {Patton}},\
  }\href {\doibase 10.1063/1.2198813} {\bibfield  {journal} {\bibinfo
  {journal} {Rev. Sci. Instrum.}\ }\textbf {\bibinfo {volume} {77}},\ \bibinfo
  {pages} {55107} (\bibinfo {year} {2006})}\BibitemShut {NoStop}%
\bibitem [{\citenamefont {Scoles}, \citenamefont {D},\ and\ \citenamefont
  {Buck}(1988)}]{Scoles1988}%
  \BibitemOpen
  \bibfield  {author} {\bibinfo {author} {\bibfnamefont {G.}~\bibnamefont
  {Scoles}}, \bibinfo {author} {\bibfnamefont {B.}~\bibnamefont {D}}, \ and\
  \bibinfo {author} {\bibfnamefont {U.}~\bibnamefont {Buck}},\ }\href@noop {}
  {\emph {\bibinfo {title} {{Atomic and Molecular Beam Methods}}}},\
  Vol.~\bibinfo {volume} {1}\ (\bibinfo  {publisher} {Oxford University Press,
  New York Oxford},\ \bibinfo {year} {1988})\ p.\ \bibinfo {pages}
  {752}\BibitemShut {NoStop}%
\bibitem [{\citenamefont {Eder}\ \emph {et~al.}(2013)\citenamefont {Eder},
  \citenamefont {Samelin}, \citenamefont {Bracco}, \citenamefont {Ansperger},\
  and\ \citenamefont {Holst}}]{Sab_freejet}%
  \BibitemOpen
  \bibfield  {author} {\bibinfo {author} {\bibfnamefont {S.~D.}\ \bibnamefont
  {Eder}}, \bibinfo {author} {\bibfnamefont {B.}~\bibnamefont {Samelin}},
  \bibinfo {author} {\bibfnamefont {G.}~\bibnamefont {Bracco}}, \bibinfo
  {author} {\bibfnamefont {K.}~\bibnamefont {Ansperger}}, \ and\ \bibinfo
  {author} {\bibfnamefont {B.}~\bibnamefont {Holst}},\ }\href@noop {}
  {\bibfield  {journal} {\bibinfo  {journal} {Rev. Sci. Instrum}\ }\textbf
  {\bibinfo {volume} {84}} (\bibinfo {year} {2013})}\BibitemShut {NoStop}%
\bibitem [{\citenamefont {Even}(2015)}]{Even2015}%
  \BibitemOpen
  \bibfield  {author} {\bibinfo {author} {\bibfnamefont {U.}~\bibnamefont
  {Even}},\ }\href {\doibase 10.1140/epjti/s40485-015-0027-5} {\bibfield
  {journal} {\bibinfo  {journal} {EPJ Tech. Instrum.}\ }\textbf {\bibinfo
  {volume} {2}},\ \bibinfo {pages} {17} (\bibinfo {year} {2015})}\BibitemShut
  {NoStop}%
\bibitem [{\citenamefont {Koch}\ \emph {et~al.}(2008)\citenamefont {Koch},
  \citenamefont {Rehbein}, \citenamefont {Schmahl}, \citenamefont {Reisinger},
  \citenamefont {Bracco}, \citenamefont {Ernst},\ and\ \citenamefont
  {Holst}}]{JMI:JMI1874}%
  \BibitemOpen
  \bibfield  {author} {\bibinfo {author} {\bibfnamefont {M.}~\bibnamefont
  {Koch}}, \bibinfo {author} {\bibfnamefont {S.}~\bibnamefont {Rehbein}},
  \bibinfo {author} {\bibfnamefont {G.}~\bibnamefont {Schmahl}}, \bibinfo
  {author} {\bibfnamefont {T.}~\bibnamefont {Reisinger}}, \bibinfo {author}
  {\bibfnamefont {G.}~\bibnamefont {Bracco}}, \bibinfo {author} {\bibfnamefont
  {W.~E.}\ \bibnamefont {Ernst}}, \ and\ \bibinfo {author} {\bibfnamefont
  {B.}~\bibnamefont {Holst}},\ }\href {\doibase
  10.1111/j.1365-2818.2007.01874.x} {\bibfield  {journal} {\bibinfo  {journal}
  {J. Microsc.}\ }\textbf {\bibinfo {volume} {229}},\ \bibinfo {pages} {1}
  (\bibinfo {year} {2008})}\BibitemShut {NoStop}%
\bibitem [{\citenamefont {Fahy}\ \emph {et~al.}(2015)\citenamefont {Fahy},
  \citenamefont {Barr}, \citenamefont {Martens},\ and\ \citenamefont
  {Dastoor}}]{FahyA_2015}%
  \BibitemOpen
  \bibfield  {author} {\bibinfo {author} {\bibfnamefont {A.}~\bibnamefont
  {Fahy}}, \bibinfo {author} {\bibfnamefont {M.}~\bibnamefont {Barr}}, \bibinfo
  {author} {\bibfnamefont {J.}~\bibnamefont {Martens}}, \ and\ \bibinfo
  {author} {\bibfnamefont {P.}~\bibnamefont {Dastoor}},\ }\href@noop {}
  {\bibfield  {journal} {\bibinfo  {journal} {Rev. Sci. Instrum}\ }\textbf
  {\bibinfo {volume} {86}},\ \bibinfo {pages} {023704} (\bibinfo {year}
  {2015})}\BibitemShut {NoStop}%
\bibitem [{\citenamefont {Eder}\ \emph {et~al.}(2012)\citenamefont {Eder},
  \citenamefont {Reisinger}, \citenamefont {Greve}, \citenamefont {Bracco},\
  and\ \citenamefont {Holst}}]{ZP_Mue_2}%
  \BibitemOpen
  \bibfield  {author} {\bibinfo {author} {\bibfnamefont {S.~D.}\ \bibnamefont
  {Eder}}, \bibinfo {author} {\bibfnamefont {T.}~\bibnamefont {Reisinger}},
  \bibinfo {author} {\bibfnamefont {M.~M.}\ \bibnamefont {Greve}}, \bibinfo
  {author} {\bibfnamefont {G.}~\bibnamefont {Bracco}}, \ and\ \bibinfo {author}
  {\bibfnamefont {B.}~\bibnamefont {Holst}},\ }\href
  {http://stacks.iop.org/1367-2630/14/i=7/a=073014} {\bibfield  {journal}
  {\bibinfo  {journal} {New. J. Phys.}\ }\textbf {\bibinfo {volume} {14}},\
  \bibinfo {pages} {73014} (\bibinfo {year} {2012})}\BibitemShut {NoStop}%
\bibitem [{\citenamefont {{Salvador Palau}}, \citenamefont {Bracco},\ and\
  \citenamefont {Holst}(2016)}]{SalvadorPalau2016}%
  \BibitemOpen
  \bibfield  {author} {\bibinfo {author} {\bibfnamefont {A.}~\bibnamefont
  {{Salvador Palau}}}, \bibinfo {author} {\bibfnamefont {G.}~\bibnamefont
  {Bracco}}, \ and\ \bibinfo {author} {\bibfnamefont {B.}~\bibnamefont
  {Holst}},\ }\href@noop {} {\bibfield  {journal} {\bibinfo  {journal} {Phys.
  Rev. A}\ } (\bibinfo {year} {2016})}\BibitemShut {NoStop}%
\bibitem [{\citenamefont {Bird}(1976)}]{Bird1976}%
  \BibitemOpen
  \bibfield  {author} {\bibinfo {author} {\bibfnamefont {G.~A.}\ \bibnamefont
  {Bird}},\ }\href {\doibase 10.1063/1.861351} {\bibfield  {journal} {\bibinfo
  {journal} {Physics of Fluids}\ }\textbf {\bibinfo {volume} {19}},\ \bibinfo
  {pages} {1486} (\bibinfo {year} {1976})}\BibitemShut {NoStop}%
\bibitem [{\citenamefont {Bird}(1994)}]{Bird1994}%
  \BibitemOpen
  \bibfield  {author} {\bibinfo {author} {\bibfnamefont {G.~a.}\ \bibnamefont
  {Bird}},\ }\href {\doibase 10.1016/0042-207X(96)80021-2} {\emph {\bibinfo
  {title} {{Molecular Gas Dynamics and Direct Simulation of Gas Flows}}}},\
  \bibinfo {edition} {1st}\ ed.\ (\bibinfo  {publisher} {Oxford University
  Press, New York Oxford},\ \bibinfo {year} {1994})\ p.\ \bibinfo {pages}
  {458}\BibitemShut {NoStop}%
\bibitem [{\citenamefont {Reisinger}\ \emph {et~al.}(2007)\citenamefont
  {Reisinger}, \citenamefont {Bracco}, \citenamefont {Rehbein}, \citenamefont
  {Schmahl}, \citenamefont {Ernst},\ and\ \citenamefont
  {Holst}}]{Reisinger2007}%
  \BibitemOpen
  \bibfield  {author} {\bibinfo {author} {\bibfnamefont {T.}~\bibnamefont
  {Reisinger}}, \bibinfo {author} {\bibfnamefont {G.}~\bibnamefont {Bracco}},
  \bibinfo {author} {\bibfnamefont {S.}~\bibnamefont {Rehbein}}, \bibinfo
  {author} {\bibfnamefont {G.}~\bibnamefont {Schmahl}}, \bibinfo {author}
  {\bibfnamefont {W.~E.}\ \bibnamefont {Ernst}}, \ and\ \bibinfo {author}
  {\bibfnamefont {B.}~\bibnamefont {Holst}},\ }\href {\doibase
  10.1021/jp076102u} {\bibfield  {journal} {\bibinfo  {journal} {J. Phys. Chem
  A}\ }\textbf {\bibinfo {volume} {111}},\ \bibinfo {pages} {12620} (\bibinfo
  {year} {2007})}\BibitemShut {NoStop}%
\bibitem [{\citenamefont {Hedgeland}\ \emph {et~al.}(2005)\citenamefont
  {Hedgeland}, \citenamefont {Jardine}, \citenamefont {Allison},\ and\
  \citenamefont {Ellis}}]{Hedgeland2005a}%
  \BibitemOpen
  \bibfield  {author} {\bibinfo {author} {\bibfnamefont {H.}~\bibnamefont
  {Hedgeland}}, \bibinfo {author} {\bibfnamefont {A.~P.}\ \bibnamefont
  {Jardine}}, \bibinfo {author} {\bibfnamefont {W.}~\bibnamefont {Allison}}, \
  and\ \bibinfo {author} {\bibfnamefont {J.}~\bibnamefont {Ellis}},\ }\href
  {\doibase 10.1063/1.2149008} {\bibfield  {journal} {\bibinfo  {journal} {Rev.
  Sci. Instrum.}\ }\textbf {\bibinfo {volume} {76}},\ \bibinfo {pages} {123111}
  (\bibinfo {year} {2005})}\BibitemShut {NoStop}%
\bibitem [{\citenamefont {Braun}\ \emph {et~al.}(1997)\citenamefont {Braun},
  \citenamefont {Day}, \citenamefont {Toennies}, \citenamefont {Witte},\ and\
  \citenamefont {Neher}}]{braun:3001}%
  \BibitemOpen
  \bibfield  {author} {\bibinfo {author} {\bibfnamefont {J.}~\bibnamefont
  {Braun}}, \bibinfo {author} {\bibfnamefont {P.~K.}\ \bibnamefont {Day}},
  \bibinfo {author} {\bibfnamefont {J.~P.}\ \bibnamefont {Toennies}}, \bibinfo
  {author} {\bibfnamefont {G.}~\bibnamefont {Witte}}, \ and\ \bibinfo {author}
  {\bibfnamefont {E.}~\bibnamefont {Neher}},\ }\href {\doibase
  10.1063/1.1148233} {\bibfield  {journal} {\bibinfo  {journal} {Rev. Sci.
  Instrum.}\ }\textbf {\bibinfo {volume} {68}},\ \bibinfo {pages} {3001}
  (\bibinfo {year} {1997})}\BibitemShut {NoStop}%
\bibitem [{\citenamefont {Verheijen}\ \emph {et~al.}(1984)\citenamefont
  {Verheijen}, \citenamefont {Beijerinck}, \citenamefont {Renes},\ and\
  \citenamefont {Verster}}]{Verheijen198463}%
  \BibitemOpen
  \bibfield  {author} {\bibinfo {author} {\bibfnamefont {M.~J.}\ \bibnamefont
  {Verheijen}}, \bibinfo {author} {\bibfnamefont {H.~C.~W.}\ \bibnamefont
  {Beijerinck}}, \bibinfo {author} {\bibfnamefont {W.~A.}\ \bibnamefont
  {Renes}}, \ and\ \bibinfo {author} {\bibfnamefont {N.~F.}\ \bibnamefont
  {Verster}},\ }\href {\doibase 10.1016/S0301-0104(84)85173-3} {\bibfield
  {journal} {\bibinfo  {journal} {Chem. Phys.}\ }\textbf {\bibinfo {volume}
  {85}},\ \bibinfo {pages} {63} (\bibinfo {year} {1984})}\BibitemShut {NoStop}%
\bibitem [{\citenamefont {Anderson}\ and\ \citenamefont
  {Fenn}(1965)}]{Anderson1965}%
  \BibitemOpen
  \bibfield  {author} {\bibinfo {author} {\bibfnamefont {J.~B.}\ \bibnamefont
  {Anderson}}\ and\ \bibinfo {author} {\bibfnamefont {J.~B.}\ \bibnamefont
  {Fenn}},\ }\href {\doibase 10.1063/1.1761320} {\bibfield  {journal} {\bibinfo
   {journal} {Physics of Fluids}\ }\textbf {\bibinfo {volume} {8}},\ \bibinfo
  {pages} {780} (\bibinfo {year} {1965})}\BibitemShut {NoStop}%
\bibitem [{\citenamefont {Beijerinck}\ and\ \citenamefont
  {Verster}(1981)}]{Beijerinck1981}%
  \BibitemOpen
  \bibfield  {author} {\bibinfo {author} {\bibfnamefont {H.~C.~W.}\
  \bibnamefont {Beijerinck}}\ and\ \bibinfo {author} {\bibfnamefont {N.~F.}\
  \bibnamefont {Verster}},\ }\href {\doibase 10.1016/0378-4363(81)90112-1}
  {\bibfield  {journal} {\bibinfo  {journal} {Physica C}\ }\textbf {\bibinfo
  {volume} {111}},\ \bibinfo {pages} {327} (\bibinfo {year}
  {1981})}\BibitemShut {NoStop}%
\bibitem [{\citenamefont {Bossel}(1974)}]{bossel1974skimming}%
  \BibitemOpen
  \bibfield  {author} {\bibinfo {author} {\bibfnamefont {U.}~\bibnamefont
  {Bossel}},\ }\href {https://books.google.no/books?id=sBLQGwAACAAJ} {\emph
  {\bibinfo {title} {{Skimming of Molecular Beams from Diverging
  Non-equilibrium Gas Jets}}}},\ Deutsche Luft- und Raumfahrt.
  Forschungsbericht\ (\bibinfo  {publisher} {Deutsche Forschungs-und
  Versuchsanstalt f{\"{u}}r Luft-und Raumfahrt},\ \bibinfo {year}
  {1974})\BibitemShut {NoStop}%
\bibitem [{\citenamefont {Sikora}(1973)}]{Sikora1973}%
  \BibitemOpen
  \bibfield  {author} {\bibinfo {author} {\bibfnamefont {G.~S.}\ \bibnamefont
  {Sikora}},\ }\emph {\bibinfo {title} {{Analysis of asymptotic behavior of
  free jets: Prediction of Molecular Beam Instensity and Velocity
  Distributions}}},\ \href@noop {} {Ph.D. thesis},\ \bibinfo  {school}
  {Princeton} (\bibinfo {year} {1973})\BibitemShut {NoStop}%
\bibitem [{\citenamefont {Witham}\ and\ \citenamefont
  {Sanchez}(2011)}]{Witham2011}%
  \BibitemOpen
  \bibfield  {author} {\bibinfo {author} {\bibfnamefont {P.}~\bibnamefont
  {Witham}}\ and\ \bibinfo {author} {\bibfnamefont {E.}~\bibnamefont
  {Sanchez}},\ }\href {\doibase 10.1063/1.3650719} {\bibfield  {journal}
  {\bibinfo  {journal} {Rev. Sci. Instrum.}\ }\textbf {\bibinfo {volume}
  {82}},\ \bibinfo {pages} {103705} (\bibinfo {year} {2011})}\BibitemShut
  {NoStop}%
\bibitem [{\citenamefont {Barr}\ \emph {et~al.}(2014)\citenamefont {Barr},
  \citenamefont {Fahy}, \citenamefont {Jardine}, \citenamefont {Ellis},
  \citenamefont {Ward}, \citenamefont {Maclaren}, \citenamefont {Allison},\
  and\ \citenamefont {Dastoor}}]{Dastoor_paper}%
  \BibitemOpen
  \bibfield  {author} {\bibinfo {author} {\bibfnamefont {M.}~\bibnamefont
  {Barr}}, \bibinfo {author} {\bibfnamefont {A.}~\bibnamefont {Fahy}}, \bibinfo
  {author} {\bibfnamefont {A.}~\bibnamefont {Jardine}}, \bibinfo {author}
  {\bibfnamefont {J.}~\bibnamefont {Ellis}}, \bibinfo {author} {\bibfnamefont
  {D.}~\bibnamefont {Ward}}, \bibinfo {author} {\bibfnamefont {D.~A.}\
  \bibnamefont {Maclaren}}, \bibinfo {author} {\bibfnamefont {W.}~\bibnamefont
  {Allison}}, \ and\ \bibinfo {author} {\bibfnamefont {P.~C.}\ \bibnamefont
  {Dastoor}},\ }\href {\doibase 10.1016/j.nimb.2014.06.028} {\bibfield
  {journal} {\bibinfo  {journal} {Nucl. Instrum. Methods}\ }\textbf {\bibinfo
  {volume} {B 340}},\ \bibinfo {pages} {76} (\bibinfo {year}
  {2014})}\BibitemShut {NoStop}%
\bibitem [{\citenamefont {Bird}(1970)}]{Bird1970}%
  \BibitemOpen
  \bibfield  {author} {\bibinfo {author} {\bibfnamefont {G.~A.}\ \bibnamefont
  {Bird}},\ }\href@noop {} {\bibfield  {journal} {\bibinfo  {journal} {AIAA
  Journal}\ }\textbf {\bibinfo {volume} {8}},\ \bibinfo {pages} {1998}
  (\bibinfo {year} {1970})}\BibitemShut {NoStop}%
\bibitem [{\citenamefont {Amirav}, \citenamefont {Even},\ and\ \citenamefont
  {Jortner}(1980)}]{asea1}%
  \BibitemOpen
  \bibfield  {author} {\bibinfo {author} {\bibfnamefont {A.}~\bibnamefont
  {Amirav}}, \bibinfo {author} {\bibfnamefont {U.}~\bibnamefont {Even}}, \ and\
  \bibinfo {author} {\bibfnamefont {J.}~\bibnamefont {Jortner}},\ }\href@noop
  {} {\bibfield  {journal} {\bibinfo  {journal} {Chem. Phys.}\ }\textbf
  {\bibinfo {volume} {51}},\ \bibinfo {pages} {31} (\bibinfo {year}
  {1980})}\BibitemShut {NoStop}%
\bibitem [{\citenamefont {Toennies}\ and\ \citenamefont
  {Winkelmann}(1977)}]{Toennies1977a}%
  \BibitemOpen
  \bibfield  {author} {\bibinfo {author} {\bibfnamefont {J.}~\bibnamefont
  {Toennies}}\ and\ \bibinfo {author} {\bibfnamefont {K.}~\bibnamefont
  {Winkelmann}},\ }\href@noop {} {\bibfield  {journal} {\bibinfo  {journal} {J.
  Chem. Phys.}\ }\textbf {\bibinfo {volume} {66}},\ \bibinfo {pages} {3965}
  (\bibinfo {year} {1977})}\BibitemShut {NoStop}%
\bibitem [{\citenamefont {Pedemonte}\ and\ \citenamefont
  {Bracco}(2003)}]{Pedemonte2003}%
  \BibitemOpen
  \bibfield  {author} {\bibinfo {author} {\bibfnamefont {L.}~\bibnamefont
  {Pedemonte}}\ and\ \bibinfo {author} {\bibfnamefont {G.}~\bibnamefont
  {Bracco}},\ }\href@noop {} {\bibfield  {journal} {\bibinfo  {journal} {J.
  Chem. Phys}\ }\textbf {\bibinfo {volume} {119}},\ \bibinfo {pages} {1433}
  (\bibinfo {year} {2003})}\BibitemShut {NoStop}%
\bibitem [{\citenamefont {Usami}\ and\ \citenamefont {Teshima}(1999)}]{Usami}%
  \BibitemOpen
  \bibfield  {author} {\bibinfo {author} {\bibfnamefont {M.}~\bibnamefont
  {Usami}}\ and\ \bibinfo {author} {\bibfnamefont {K.}~\bibnamefont
  {Teshima}},\ }\href@noop {} {\bibfield  {journal} {\bibinfo  {journal} {JSME
  Int. J}\ }\textbf {\bibinfo {volume} {42}},\ \bibinfo {pages} {369} (\bibinfo
  {year} {1999})}\BibitemShut {NoStop}%
\bibitem [{\citenamefont {Holst}(2015)}]{Holst2015}%
  \BibitemOpen
  \bibfield  {author} {\bibinfo {author} {\bibfnamefont {B.}~\bibnamefont
  {Holst}},\ }\href@noop {} {\bibfield  {journal} {\bibinfo  {journal} {Rev.
  Sci. Instrum.}\ ,\ \bibinfo {pages} {1}} (\bibinfo {year}
  {2015})}\BibitemShut {NoStop}%
\bibitem [{\citenamefont {Pauly}(2000{\natexlab{b}})}]{pauly2000atom}%
  \BibitemOpen
  \bibfield  {author} {\bibinfo {author} {\bibfnamefont {H.}~\bibnamefont
  {Pauly}},\ }\href {https://books.google.no/books?id=Lwh1uDMAwDgC} {\emph
  {\bibinfo {title} {{Atom, Molecule, and Cluster Beams I: Basic Theory,
  Production and Detection of Thermal Energy Beams}}}},\ Atom, Molecule, and
  Cluster Beams\ (\bibinfo  {publisher} {Springer},\ \bibinfo {year}
  {2000})\BibitemShut {NoStop}%
\bibitem [{\citenamefont {Holst}\ \emph {et~al.}()\citenamefont {Holst},
  \citenamefont {Eder}, \citenamefont {{Salvador Palau}},\ and\ \citenamefont
  {Bracco}}]{Holst}%
  \BibitemOpen
  \bibfield  {author} {\bibinfo {author} {\bibfnamefont {B.}~\bibnamefont
  {Holst}}, \bibinfo {author} {\bibfnamefont {S.}~\bibnamefont {Eder}},
  \bibinfo {author} {\bibfnamefont {A.}~\bibnamefont {{Salvador Palau}}}, \
  and\ \bibinfo {author} {\bibfnamefont {G.}~\bibnamefont {Bracco}},\
  }\href@noop {} {\bibinfo  {journal} {Review of Scientific Instruments (under
  review)}\ }\BibitemShut {NoStop}%
\bibitem [{\citenamefont {Jones}(1924)}]{Jones1924}%
  \BibitemOpen
\bibfield  {journal} {  }\bibfield  {author} {\bibinfo {author} {\bibfnamefont
  {J.~E.}\ \bibnamefont {Jones}},\ }\href {\doibase 10.1098/rspa.1924.0081}
  {\bibfield  {journal} {\bibinfo  {journal} {Proc. R. Soc. A}\ }\textbf
  {\bibinfo {volume} {106}},\ \bibinfo {pages} {441} (\bibinfo {year}
  {1924})}\BibitemShut {NoStop}%
\bibitem [{\citenamefont {Longo}\ \emph {et~al.}(2008)\citenamefont {Longo},
  \citenamefont {Diomede}, \citenamefont {Laricchiuta}, \citenamefont
  {Colonna}, \citenamefont {Capitelli}, \citenamefont {Ascenzi}, \citenamefont
  {Scoltoni}, \citenamefont {Tosi},\ and\ \citenamefont
  {Pirani}}]{Proceedingspotential}%
  \BibitemOpen
  \bibfield  {author} {\bibinfo {author} {\bibfnamefont {S.}~\bibnamefont
  {Longo}}, \bibinfo {author} {\bibfnamefont {P.}~\bibnamefont {Diomede}},
  \bibinfo {author} {\bibfnamefont {A.}~\bibnamefont {Laricchiuta}}, \bibinfo
  {author} {\bibfnamefont {G.}~\bibnamefont {Colonna}}, \bibinfo {author}
  {\bibfnamefont {G.}~\bibnamefont {Capitelli}}, \bibinfo {author}
  {\bibfnamefont {D.}~\bibnamefont {Ascenzi}}, \bibinfo {author} {\bibfnamefont
  {M.}~\bibnamefont {Scoltoni}}, \bibinfo {author} {\bibfnamefont
  {P.}~\bibnamefont {Tosi}}, \ and\ \bibinfo {author} {\bibfnamefont
  {F.}~\bibnamefont {Pirani}},\ }in\ \href@noop {} {\emph {\bibinfo {booktitle}
  {Lecture Notes in Computer Science 2008}}},\ \bibinfo {editor} {edited by\
  \bibinfo {editor} {\bibfnamefont {G.}~\bibnamefont {et~al}}}\ (\bibinfo
  {publisher} {Springer-Verlag},\ \bibinfo {address} {Berlin},\ \bibinfo {year}
  {2008})\ p.\ \bibinfo {pages} {1131}\BibitemShut {NoStop}%
\bibitem [{\citenamefont {Tang}, \citenamefont {Toennies},\ and\ \citenamefont
  {Yiu}(1995)}]{PhysRevLett.74.1546}%
  \BibitemOpen
  \bibfield  {author} {\bibinfo {author} {\bibfnamefont {K.~T.}\ \bibnamefont
  {Tang}}, \bibinfo {author} {\bibfnamefont {J.~P.}\ \bibnamefont {Toennies}},
  \ and\ \bibinfo {author} {\bibfnamefont {C.~L.}\ \bibnamefont {Yiu}},\ }\href
  {\doibase 10.1103/PhysRevLett.74.1546} {\bibfield  {journal} {\bibinfo
  {journal} {Phys. Rev. Lett.}\ }\textbf {\bibinfo {volume} {74}},\ \bibinfo
  {pages} {1546} (\bibinfo {year} {1995})}\BibitemShut {NoStop}%
\bibitem [{\citenamefont {Hurly}\ and\ \citenamefont
  {Moldover}(2000)}]{Hurly2000}%
  \BibitemOpen
  \bibfield  {author} {\bibinfo {author} {\bibfnamefont {J.~J.}\ \bibnamefont
  {Hurly}}\ and\ \bibinfo {author} {\bibfnamefont {M.~R.}\ \bibnamefont
  {Moldover}},\ }\href@noop {} {\bibfield  {journal} {\bibinfo  {journal} {J.
  Res. Natl. Inst. Stand. Technol.}\ }\textbf {\bibinfo {volume} {105}},\
  \bibinfo {pages} {667} (\bibinfo {year} {2000})}\BibitemShut {NoStop}%
\bibitem [{\citenamefont {Pedemonte}, \citenamefont {Bracco},\ and\
  \citenamefont {Tatarek}(1999)}]{PhysRevA.59.3084}%
  \BibitemOpen
  \bibfield  {author} {\bibinfo {author} {\bibfnamefont {L.}~\bibnamefont
  {Pedemonte}}, \bibinfo {author} {\bibfnamefont {G.}~\bibnamefont {Bracco}}, \
  and\ \bibinfo {author} {\bibfnamefont {R.}~\bibnamefont {Tatarek}},\ }\href
  {\doibase 10.1103/PhysRevA.59.3084} {\bibfield  {journal} {\bibinfo
  {journal} {Phys. Rev. A}\ }\textbf {\bibinfo {volume} {59}},\ \bibinfo
  {pages} {3084} (\bibinfo {year} {1999})}\BibitemShut {NoStop}%
\bibitem [{\citenamefont {{Salvador Palau}}\ \emph {et~al.}(2015)\citenamefont
  {{Salvador Palau}}, \citenamefont {Eder}, \citenamefont {Kaltenbacher},
  \citenamefont {Samelin}, \citenamefont {Bracco},\ and\ \citenamefont
  {Holst}}]{SalvadorPalau2015}%
  \BibitemOpen
  \bibfield  {author} {\bibinfo {author} {\bibfnamefont {A.}~\bibnamefont
  {{Salvador Palau}}}, \bibinfo {author} {\bibfnamefont {S.~D.}\ \bibnamefont
  {Eder}}, \bibinfo {author} {\bibfnamefont {T.}~\bibnamefont {Kaltenbacher}},
  \bibinfo {author} {\bibfnamefont {B.}~\bibnamefont {Samelin}}, \bibinfo
  {author} {\bibfnamefont {G.}~\bibnamefont {Bracco}}, \ and\ \bibinfo {author}
  {\bibfnamefont {B.}~\bibnamefont {Holst}},\ }\href@noop {} {\bibfield
  {journal} {\bibinfo  {journal} {Rev. Sci. Instrum.}\ } (\bibinfo {year}
  {2015})}\BibitemShut {NoStop}%
\bibitem [{\citenamefont {{Salvador Palau}}, \citenamefont {Bracco},\ and\
  \citenamefont {Holst}(2017)}]{PhysRevA.95.013611_ZP}%
  \BibitemOpen
  \bibfield  {author} {\bibinfo {author} {\bibfnamefont {A.}~\bibnamefont
  {{Salvador Palau}}}, \bibinfo {author} {\bibfnamefont {G.}~\bibnamefont
  {Bracco}}, \ and\ \bibinfo {author} {\bibfnamefont {B.}~\bibnamefont
  {Holst}},\ }\href {\doibase 10.1103/PhysRevA.95.013611} {\bibfield  {journal}
  {\bibinfo  {journal} {Phys. Rev. A}\ }\textbf {\bibinfo {volume} {95}},\
  \bibinfo {pages} {13611} (\bibinfo {year} {2017})}\BibitemShut {NoStop}%
\bibitem [{\citenamefont {Luria}, \citenamefont {Christen},\ and\ \citenamefont
  {Even}(2011)}]{Luria2011}%
  \BibitemOpen
  \bibfield  {author} {\bibinfo {author} {\bibfnamefont {K.}~\bibnamefont
  {Luria}}, \bibinfo {author} {\bibfnamefont {W.}~\bibnamefont {Christen}}, \
  and\ \bibinfo {author} {\bibfnamefont {U.}~\bibnamefont {Even}},\ }\href
  {http://pubs.acs.org/doi/abs/10.1021/jp201342u} {\bibfield  {journal}
  {\bibinfo  {journal} {J. Phys. Chem A}\ }\textbf {\bibinfo {volume} {115}},\
  \bibinfo {pages} {7362} (\bibinfo {year} {2011})}\BibitemShut {NoStop}%
\bibitem [{\citenamefont {Lefmann}\ and\ \citenamefont
  {Nielsen}(1999)}]{doi:10.1080/10448639908233684}%
  \BibitemOpen
  \bibfield  {author} {\bibinfo {author} {\bibfnamefont {K.}~\bibnamefont
  {Lefmann}}\ and\ \bibinfo {author} {\bibfnamefont {K.}~\bibnamefont
  {Nielsen}},\ }\href {\doibase 10.1080/10448639908233684} {\bibfield
  {journal} {\bibinfo  {journal} {Neutron News}\ }\textbf {\bibinfo {volume}
  {10}},\ \bibinfo {pages} {20} (\bibinfo {year} {1999})}\BibitemShut {NoStop}%
\bibitem [{\citenamefont {Willendrup}, \citenamefont {Farhi},\ and\
  \citenamefont {Lefmann}(2004)}]{Willendrup2004}%
  \BibitemOpen
  \bibfield  {author} {\bibinfo {author} {\bibfnamefont {P.}~\bibnamefont
  {Willendrup}}, \bibinfo {author} {\bibfnamefont {E.}~\bibnamefont {Farhi}}, \
  and\ \bibinfo {author} {\bibfnamefont {K.}~\bibnamefont {Lefmann}},\ }in\
  \href {\doibase 10.1016/j.physb.2004.03.193} {\emph {\bibinfo {booktitle}
  {Physica. B. Condens. Matter}}},\ Vol.\ \bibinfo {volume} {350}\ (\bibinfo
  {year} {2004})\BibitemShut {NoStop}%
\bibitem [{\citenamefont {Apfolter}(2005)}]{Apfolter2005}%
  \BibitemOpen
  \bibfield  {author} {\bibinfo {author} {\bibfnamefont {A.}~\bibnamefont
  {Apfolter}},\ }\href@noop {} {\enquote {\bibinfo {title} {{Master Thesis}},}\
  } (\bibinfo {year} {2005})\BibitemShut {NoStop}%
\bibitem [{\citenamefont {Dynamics}()}]{Beamdynamics}%
  \BibitemOpen
  \bibfield  {author} {\bibinfo {author} {\bibfnamefont {B.}~\bibnamefont
  {Dynamics}},\ }\href@noop {} {\bibinfo  {journal} {Beam Dynamics}\
  }\BibitemShut {NoStop}%
\bibitem [{\citenamefont {Doak}\ \emph {et~al.}(1999)\citenamefont {Doak},
  \citenamefont {Grisenti}, \citenamefont {Rehbein}, \citenamefont {Schmahl},
  \citenamefont {Toennies},\ and\ \citenamefont
  {W{\"{o}}ll}}]{PhysRevLett_4229}%
  \BibitemOpen
\bibfield  {journal} {  }\bibfield  {author} {\bibinfo {author} {\bibfnamefont
  {R.~B.}\ \bibnamefont {Doak}}, \bibinfo {author} {\bibfnamefont {R.~E.}\
  \bibnamefont {Grisenti}}, \bibinfo {author} {\bibfnamefont {S.}~\bibnamefont
  {Rehbein}}, \bibinfo {author} {\bibfnamefont {G.}~\bibnamefont {Schmahl}},
  \bibinfo {author} {\bibfnamefont {J.~P.}\ \bibnamefont {Toennies}}, \ and\
  \bibinfo {author} {\bibfnamefont {C.}~\bibnamefont {W{\"{o}}ll}},\ }\href
  {\doibase 10.1103/PhysRevLett.83.4229} {\bibfield  {journal} {\bibinfo
  {journal} {Phys. Rev. Lett.}\ }\textbf {\bibinfo {volume} {83}},\ \bibinfo
  {pages} {4229} (\bibinfo {year} {1999})}\BibitemShut {NoStop}%
\end{thebibliography}

\end{document}